\documentclass[12pt,dvipdfmx]{iopart}

\usepackage{amssymb}
\usepackage{bm}
\usepackage{color}
\usepackage[english]{babel}
\usepackage[dvipdfmx]{graphicx}

\begin{document}

\title[]
{A new equation of state for core-collapse supernovae based on realistic nuclear forces and including a full nuclear ensemble}

\author{S.~Furusawa$^{1,2}$, H.~Togashi$^{3,4}$,  H.~Nagakura$^{5}$, K.~Sumiyoshi$^{6}$, S.~Yamada$^{7}$, H.~Suzuki$^{8}$ and M.~Takano$^{4,7}$}
\address{$^1$ Interdisciplinary Theoretical Science (iTHES) Research Group, RIKEN, Wako, Saitama 351-0198, Japan}
\address{$^2$Frankfurt Institute for Advanced Studies, J.W. Goethe University, 60438 Frankfurt am Main, Germany}
\address{$^3$ RIKEN Nishina Center, RIKEN, Saitama 351-0198, Japan}
\address{$^4$ Research Institute for Science and Engineering, Waseda University, Tokyo 169-8555, Japan}
\address{$^5$ TAPIR, Walter Burke Institute for Theoretical Physics, Mailcode 350-17, California Institute of Technology, Pasadena, CA 91125, USA}
\address{$^6$ Numazu College of Technology, Ooka 3600, Numazu, Shizuoka 410-8501, Japan}
\address{$^7$ Department of Science and Engineering, Waseda University, 3-4-1 Okubo, Shinjuku, Tokyo 169-8555, Japan}
\address{$^8$ Faculty of Science and Technology, Tokyo University of Science, Yamazaki 2641, Noda, Chiba 278-8510, Japan}

\vspace{10pt}
\begin{indented}
\item\today
\end{indented}

\begin{abstract}
We have constructed a nuclear equation of state (EOS) that includes
a full nuclear ensemble for use in core-collapse supernova simulations. It is based on the EOS for uniform nuclear matter that two of the authors derived recently, applying 
a variational method to realistic two- and there-body nuclear forces. 
We have extended the  liquid drop model of heavy nuclei, utilizing the mass formula that  accounts for the dependences of bulk, surface, Coulomb  and shell  energies on density and/or temperature.
As for light nuclei,  we employ a quantum-theoretical mass evaluation,
which incorporates the Pauli- and self-energy shifts.
In addition to realistic nuclear forces,  the inclusion of in-medium effects on the full ensemble of nuclei makes
the new EOS  one of  the most realistic  EOS's, which covers 
a wide range of density, temperature and proton fraction that 
 supernova simulations normally encounter. 
We make comparisons with the FYSS  EOS, which is based on the same formulation for the nuclear ensemble  but  adopts the relativistic mean field  (RMF)  theory with the TM1 parameter set for uniform nuclear matter.
The new EOS is softer than the FYSS EOS around and above nuclear saturation densities. 
We find that 
 neutron-rich nuclei with small mass numbers 
are more abundant in the new EOS
than in  the  FYSS EOS because of the  larger saturation densities and smaller  symmetry energy of nuclei in the former.
We
apply
the two EOS's to 1D supernova simulations
and  find that
the new EOS gives lower electron fractions and higher temperatures in the collapse phase owing to the smaller symmetry energy. As a result, the inner core has smaller masses for the new EOS. It is more compact, on the other hand, due to the softness of the new EOS and bounces at higher densities. 
It turns out that the shock wave generated by core bounce is a bit stronger initially in the simulation with the new EOS. 
The ensuing outward propagations of the shock wave in the outer core are very similar in the two simulations, which may be an artifact, though,  caused by the use of the same tabulated electron capture rates for heavy nuclei  ignoring differences in the nuclear composition between the two EOS's
in these computations. 
\end{abstract}

\section{Introduction \label{intro}}
The nuclear equation of state (EOS) is an important ingredient 
to determine the dynamics of  core collapse supernovae 
\cite{janka12,kotake12,burrows13,foglizzo15}. 
It provides information on  thermodynamical quantities such as pressure and composition of high-density matter realized in the supernova core, 
which is either  a mixture of nuclei or  uniformly distributed nucleons, for given baryon number density $n_B$, temperature $T$ and proton fraction $Y_p$.
The pressure is obviously a quantity that has a direct influence on
 the dynamics  of supernova,  determining, e.g.,  the strength of the shock wave at core bounce
and the evolution of proto-neutron stars (PNS) in the post-bounce phase
\cite{sumiyoshi05,marek09,suwa13,fischer13,togashi14}.
The  abundance of various nuclei is important, affecting the neutronization of matter through electron captures on nuclei and the opacity for neutrino via the coherent scattering on heavy nuclei \cite{hix03,lentz12,sullivan16}. 
Note that the population of light nuclei such as deuteron, triton, helion and alpha particles may have some implications for shock revival \cite{sumiyoshi08,furusawa13b,nasu15,fischer16}.

One of the critical elements in the construction of the nuclear EOS is the theoretical approach 
to nuclear interactions, on which
 the bulk properties of individual nuclei as well as  the free energy of homogeneous nuclear matter depend strongly.
In the previous EOS's used in supernova simulations,
Skyrme type interactions \cite{lattimer91}  or 
meson-exchange models were adopted
\cite{shen98a,shen98b,shen11,hempel10,steiner13,sheng11,furusawa11,furusawa13a,furusawa17a}.
Both of them are based not on realistic interactions in vacuum
that are known experimentally
 but on phenomenological medium-dependent interactions, whose model parameters are 
determined so that some nuclear properties
 such as the saturation density are reproduced.
Recently, Togashi et al.  \cite{togashi17}  have constructed an EOS based on a realistic two-body nuclear potential,  Argonne v18  \cite{wiringa95},
 employing the variational method \cite{togashi13}.
This is the first EOS of such kind, meant for supernova simulations.


The calculation of the  EOS for homogeneous nuclear matter 
is not sufficient if the EOS were to be used in supernova simulations.
We need in addition to treat inhomogeneous nuclear matter composed of various nuclei and dripped nucleons,
 which exists generically at sub-nuclear densities.
The common choice is 
the so-called single nucleus approximation (SNA), in which
 the ensemble of heavy nuclei is represented just by a single, supposedly the most abundant nucleus.
It was  adopted in the two  standard EOSs, widely employed in the simulations of core-collapse  supernovae \cite{lattimer91,shen98a,shen98b,shen11} and also in the latest EOS  \cite{togashi17}. 
Either  the compressible liquid drop model (LDM)  or the Thomas-Fermi 
approximation was employed in those EOS's. 
In such approximations, 
 in-medium effects on the representative  nucleus, such as compression, can be taken into account easily. 
Note that the abundance of various nuclei is important for the
 estimation of
neutralization rates \cite{furusawa17b}
and that it should be determined consistently with the modifications of nuclear properties in medium, since nuclear masses depends on the population of other nuclei whereas the nuclear abundance is dictated by the nuclear masses.
It was demonstrated that  the average mass number and the total mass fraction of heavy nuclei 
do not coincide with those of 
 the representative nucleus in general \cite{burrows84, furusawa17c}.

In the last few years, some research groups
have constructed the multi-nuclear species EOS's \cite{botvina04, botvina10,hempel10,buyukcizmeci14},
 in which the full ensemble of nuclei are 
taken into account with 
 various finite-density and -temperature effects, such as the formation
of nuclear pastas \cite{watanabe05,newton09,okamoto12,schneider13,horowitz16},
 smearing of shell effects in heavy nuclei \cite{brack74, bohr87, sandulescu97,nishimura14} 
and self- and Pauli-energy shifts in light nuclei  \cite{typel10,roepke09},
being only partially considered. 
We have also constructed 
such an EOS  (FYSS EOS) 
\cite{furusawa11,furusawa13a, furusawa17a},
in which  the mass formula was extended for heavy nuclei 
and took
a quantum approach for light clusters 
to accommodate all the in-medium effects mentioned above.
Although the FYSS EOS may be better in the treatment of in-medium effects, it is sometimes pointed out that the  relativistic mean field (RMF) with the TM1 parameter set \cite{sugahara94}, on which FYSS EOS is based, gives too high a symmetry energy ~36.9 MeV compared with the conventional values of ~29.0-33.7 MeV \cite{lattimer13,danielewicz14,oertel17}.


The purpose of the present study is to 
construct a new, multi-nuclear species  EOS
in the framework  of the FYSS EOS
but with a different EOS for uniform nuclear matter as a base.
 We use the EOS by Togashi et al. \cite{togashi13}, who applied the variational method, one of the most popular techniques for the construction of the EOS of cold neutron star matter, using realistic nuclear forces expressed with the Argonne v18 potential \cite{wiringa95}. Note that in their EOS the symmetry energy is ~30.0 MeV,  and with the inclusion of three-body forces, the saturation properties of uniform nuclear matter are satisfactorily reproduced. 
This is the first multi-nucleus EOS based on realistic nuclear forces. 
This article is organized as follows.
In section~\ref{sec:model}, we describe the 
formulation adopted in this paper to construct  the EOS.
The results  are shown in section~\ref{sec:res}
in comparison with  the FYSS EOS derived from the RMF with the TM1 parameter set.
We also  compare the applications of these two EOS's to the one-dimensional supernova simulation in section~\ref{sec:simu}.
The paper is wrapped up with a summary and some discussion in section~\ref{sec:conc}.

\section{Formulation \label{sec:model}}
The calculation  below is based  on the formulation of the FYSS EOS 
 to describe the nuclear ensemble  \cite{furusawa11,furusawa13a,furusawa17a} 
and the variational principle method for uniform nuclear matter 
\cite{togashi13,togashi17}.
 For details, we refer the reader to  Furusawa et al.  \cite{furusawa17a} and Togashi et al. \cite{togashi13}.
The free energy of the  FYSS EOS is represented  as
\begin{eqnarray}
\label{eq:total}
\ f = f_{p,n}+\sum_{j}{n_j (E^t_{j} + M_{j}) } + \sum_i {n_i (E^t_{i} + M_{i})}, 
\end{eqnarray}
where $f_{p,n}$ is the free energy density of the free nucleons outside nuclei, $n_{j/i}$ are the number densities of individual light nuclei $j$   with the proton number  $Z_j\leq 5$
and heavy nuclei $i$  with the proton number $6 \leq Z_i \leq 1000$,  $E^t_{i/j}$ are 
the translational energies of heavy and light nuclei
and $M_{i/j}$ are masses of heavy and light nuclei.
In this work,  the variational calculation takes the place of  the RMF calculation in the previous model of the FYSS EOS, which provides the free energy density of nucleons $f_{p,n}$, the bulk energies of heavy nuclei in $M_i$ and the self-energy shifts of light nuclei in $M_j$.

\subsection{Free energy of free nucleons}
The variational calculations for the free energy of free nucleons 
  are based on Refs. \cite{togashi13,kanzawa07,kanzawa09}.
We start from the nuclear Hamiltonian composed of a two-body potential $V_{ij}$ and three-body potentials $V_{ijk}$, as in the Fermi Hypernetted Chain  variational calculations \cite{akmal98}:
\begin{eqnarray}
\label{VM}
H=-\sum^N_{i=1}  \frac{\hbar^2}{2 m} \nabla^2 +\sum^N_{i<j}V_{ij} + \sum_{i<j<k}^{N} V_{ijk},
\end{eqnarray} 
where $m$ is set to be the mass of  neutron. 

The free energy derived from two body interactions is obtained by using 
an extension of the variational method by Schmidt and Pandharipande \cite{schmidt79, mukherjee07} with AV18 two-body potential \cite{wiringa95} and the healing distance
condition, which reproduces the internal energy per baryon of symmetric
nuclear matter and neutron matter at zero temperature obtained by Akmal et al. \cite{akmal98}.
The internal energy for three-body interactions is
based on  the UIX three-body potential  \cite{carlson83,pudliner95} 
with a phenomenological correction  that reproduces  empirical saturation properties. 
The entropy is expressed with average occupation probabilities of single quasi-nucleon states as in the case of a non-interacting Fermi gas
 \cite{schmidt79}.  
The total free energy per baryon is minimized with respect to effective mases for nucleons. 
The optimized free energies agree reasonably with those of  Akmal et al. \cite{akmal98} at zero temperature
and those of Mukherjee  \cite{mukherjee09} at finite temperature. 

We account for the excluded-volume effect. 
The local number densities of free protons and neutrons are defined as $n'_{p/n}=(N_{p/n})/(V-V_N)$  where $V$ is the total volume, $V_N$ is the volume of all nuclei, and  $N_{p/n}$ are the numbers of free protons and  neutrons.
The free energy density  of free nucleons is given by 
\begin{equation}
\label{fpn}
f_{p,n} =\frac{(V-V_N)}{V}   (n_p'+n'_n)   \omega(T,n'_p+n'_n,n'_p/(n_p'+n'_n)),
\end{equation}
where $\omega(T,n_B,x)$ is the free energy density per baryon in the unoccupied volume, $V-V_N$, for nucleons obtained from the variational calculation at temperature $T$, the local number density of baryon $n_B=n'_p+n'_n$  and the charge fraction $x=n'_p/(n_p'+n'_n)$.

\subsection{Masses of heavy nuclei}
The masses of heavy nuclei are assumed to be the sum of the  bulk, Coulomb, surface  and shell energies: $M_i=E_i^B+E_i^C+E_i^{Sf} +E_i^{Sh}$.
We define the saturation density of nuclei $n_{si}(T)$ 
as the baryon number density, at which the free energy per baryon
 $\omega(T,n_B, Z_i/A_i)$ given by the variational method  becomes a minimum, 
which is the same as that for the free energy density of free nucleons. 
Thus $n_{si} (T)$ depends on the temperature $T$ and the charge fraction in each nucleus $Z_i/A_i$.
At high temperatures $T \geq T_{ci}$, the free energy, $\omega(T,n_B, Z_i/A_i)$, has no minimum because of finite entropy. 
Then the saturation density $n_{si} (T)$ above $T_{ci}$  is  assumed to be equal to 
 $n_{si}(T_{ci})$. 
When the saturation density $n_{si}$ so obtained  is lower than the baryon number density of the whole system $n_B$, we reset the saturation density as the baryon number density $n_{si}=n_B$.
This treatment of the saturation density is needed to  obtain reasonable bulk energies at high temperatures and densities \cite{furusawa13a}. 

The bulk,  surface, Coulomb and shell terms in $M_i$ are expressed as
\begin{eqnarray}
E_{i}^{B}=  A_i \{\omega(T,n_{si},Z_i/A_i) \},  \label{eq:bulk} \\
E_i^C=\left\{ \begin{array}{ll}
\displaystyle{\frac{3}{5}\left(\frac{3}{4 \pi}\right)^{-1/3}  e^2 n_{si}^2 \left(\frac{Z_i - n'_p V_i^N}{A_i}\right)^2 {V_i^N}^{5/3} D(u_i)}    & (u_i\leq 0.3), \\
\displaystyle{\frac{3}{5}\left(\frac{3}{4 \pi}\right)^{-1/3}   e^2 n_{si}^2 \left(\frac{Z_i - n'_p V_i^N}{A_i}\right)^2 {V_i^B}^{5/3} D(1-u_i)}   & (u_i\geq 0.7),
\end{array} \right. \label{eqclen}  \\
E_i^{Sf}=\left\{ \begin{array}{ll}
4 \pi {r^2_{Ni}} \, \sigma_i \left(1-\displaystyle{\frac{n'_p+n'_n}{n_{si}}} \right)^2 \left( \displaystyle{\frac{T^2_{cs}-T^2}{T^2_{cs}+T^2}} \right)^{5/4}   & (u_i\leq 0.3), \\
4 \pi {r^2_{Bi}} \, \sigma_i  \left(1-\displaystyle{\frac{n'_p+n'_n}{n_{si}}} \right)^2 \left( \displaystyle{\frac{T^2_{cs}-T^2}{T^2_{cs}+T^2}} \right)^{5/4}  & (u_i\geq 0.7), 
\end{array} \right. \label{eq:surf} \\
 E_{i}^{Sh}=\left\{ \begin{array}{ll}
E_{i0}^{Sh} \displaystyle{\frac{\tau_i}{{\rm sinh}\tau_i}} & (\rho \leq 10^{12} \rm{g/cm^3}), \\
\\
E_{i0}^{Sh} \displaystyle{\left( \frac{\tau_i}{{\rm sinh}\tau_i}\right) \left( \frac{\rho_0- \rho}{\rho_0 - 10^{12}  \rm{g/cm^3}} \right) }& (\rho >  10^{12} \rm{g/cm^3}).
\end{array} \right.    \label{eq:shell} 
\end{eqnarray} 
The bulk energy $E_{i}^{B}$ is evaluated from the free energy per baryon of uniform nuclear matter  
$\omega(T,n_B,x)  $ given by the variational method
 at   $n_B=n_{si}$ and   $x=Z_i/A_i$ for a given temperature $T$.

The Coulomb  energy $E_i^C$ is 
obtained  by integrating  the Coulomb potential over the Wigner-Seitz cell  containing nucleus $i$, dripped free protons and uniformly  distributed   electrons \cite{furusawa11}.
The cell is set to satisfy charge neutrality 
 within the volume
 as $V_i = (Z_i - n'_p  V_{Ni})/(n_e-n'_p)$, 
where $V_{Ni}$ is the volume  of the nucleus in the cell and can be calculated as $V_{Ni} = A_i / n_{si}$ and 
 $n_e=Y_p n_B$  is  the number density of electrons. 
The vapor volume and nucleus volume fraction in the cell are given by $ V_i^B = V_i-V_i^N  $ and $u_i =  V_i^N / V_i$, respectively.
Then $E_i^C$ is  expressed  as Eq.~(\ref{eqclen}) where $D(u_i)=1-\frac{3}{2}u_i^{1/3}+\frac{1}{2}u_i$ and  $e$ is the elementary charge.
Each nucleus is assumed to enter the nuclear pasta phase individually when 
the volume fraction reaches $u_i=0.3$ and that the bubble shape is realized when it exceeds $0.7$. 

The surface energy $E_i^{Sf}$ is obtained from surface area times surface tension 
 $\sigma_{i}=\sigma_0  - A_i^{2/3}/(4 \pi r_{Ni}^2) [S_s(1- 2(Z_i/A_i)^2) ]$, where 
the values of the constants, $\sigma_0=1.15$ MeV/fm$^3$ and 
$S_s =$ 45.8 MeV, are adopted from  Lattimer and Swesty \cite{lattimer91}. 
We may have to choose the values that are consistent with the EOS’s for uniform nuclear matter. 
The simpler estimate, however, is adopted in this work, since there is no concrete way to derive the surface coefficients only from the EOS for uniform nuclear matter. 
The last two factors  in Eq.~(\ref{eq:surf}) 
describe the reduction of the surface energy 
 when the density contrast between the nucleus and the nucleon vapor  decreases  and when the temperature grows.
The critical temperature is set to be  $T_{cs}=18$ MeV \cite{bondorf95} and 
the radii of nuclei and bubbles are expressed as $r_{Ni} = ( 3/4 \pi V^N_i)^{1/3}$ and $r_{Bi} = ( 3/4 \pi V^B_i)^{1/3}$.
We use cubic polynomials of $u_i$ for the interpolation of Coulomb and surface energies  between the droplet  and bubble phases  to ensure continuous and smooth connections at $u_i=0.3$ and $u_i=0.7$. 
We referred Watanabe et al. \cite{watanabe05} for the criterion
of intermediate states, although the thermodynamic
quantities are hardly affected by its choice.
 The interpolation  is mainly required to ensure the smooth change in mass fractions of nuclei  \cite{furusawa13a}. 

The shell energy at zero density and temperature  is assumed to be the difference between the experimental or theoretical mass data \cite{audi12,koura05}  and our liquid drop mass formula in the vacuum limit  without the shell term: $E_{i0}^{Sh} =M_i^{data} -[E_i^B+E_i^C+E_i^{Sf}]_{T, n'_{p/n}, n_e =0}$.
The  factor $\tau_i/{\rm sinh}\tau_i$ in  Eq.~(\ref{eq:shell})  expresses the washout of shell effects approximately with  $\tau_i =2 \pi^2 T/(41 A_i^{-1/3})$, which 
 reproduces qualitatively the feature that shell effects disappear around $T\sim$~2.0－3.0 MeV  \cite{furusawa17a}.
The linear interpolation, $(\rho_0- \rho )/(\rho_0 - 10^{12} \rm{g/cm^3})$, accounts for the disappearance of shell effects at high densities where $\rho$ and $\rho_0$  are the density and  the saturation density, respectively  \cite{furusawa11}. 

\subsection{Masses of light nuclei}
We describe 
light nuclei 
as  quasi-particles outside heavy nuclei, whose masses  are assumed to be given by  
  $ M_{j}=M_j^{data} + \Delta E_{j}^{Pa}  + \Delta E_j^{SE} +\Delta E_j^C$. 
 \label{eqquasi}
The  Pauli-energy shift $\Delta E_j^{Pa}$ 
for deuteron  ($d$), triton  ($t$), helion ($h$) and alpha particle  ($\alpha$),
which is fitted to the result of quantum statistical calculations \cite{roepke09}, 
 is given by \cite{typel10}
\begin{eqnarray}
\label{DEP}
 \Delta E_{j}^{Pa}(n_{pl},n_{nl},T)  =  - \tilde{n}_{j} 
 \left[1 + \frac{\tilde{n}_{j}}{2\tilde{n}_{j}^{0}(T)} \right]
 \delta B_{j} (T), \\
 \delta B_{j}(T) =\left\{ \begin{array}{ll}
  a_{j,1} / T^{3/2}  \left[ 1/\sqrt{y_{j}}  -  \sqrt{\pi}a_{j,3} \exp  \left( a_{j,3}^{2} y_{j} \right) 
 {\rm erfc} \left(a_{j,3} \sqrt{y_{j}} \right) \right]   &  {\rm{for}} \   j  =  d, \\
  a_{j,1} /\left( T  y_j \right)^{3/2}  \ \   &  {\rm{for}} \  j  =  t,h,\alpha, 
\end{array} \right.
\end{eqnarray}
where  $\tilde{n}_{j} = 2(Z_{j} \ n_{pl} +N_{j} \ n_{nl}) /A_{j}$, $y_{j} = 1+a_{j,2}/T$, 
 $n_{pl/nl}$ are the local proton and neutron number densities that include light nuclei as well as free nucleons
and   $\tilde{n}_{j}^{0}(T) = B_{j}^{0}/\delta B_{j}(T)$ with the binding energy in vacuum, $B_{j}^{0}$. 
The Pauli-energy shifts for  the light nuclei other than  $d, t, h$ and $\alpha$  are calculated 
 in the same way as that  for  $\alpha$. 

The self-energy shift  $\Delta E_j^{SE}$ is the sum of
 the self-energy shifts of individual nucleons in light nuclei 
 $\Delta E^{SE}_{n/p}=\Sigma^0_{n/p}(T,n'_p,n'_n)-\Sigma_{n/p}(T,n'_p,n'_n)$ with
  $\Sigma^0$ and $\Sigma$ being the vector and scalar potentials of nucleon and 
the contribution from their effective masses $\Delta E_{j}^{\rm eff.mass} = \left(1-m^{\ast}/m \right)s_{j}$ with $m^{\ast}=m_B-\Sigma_{n/p}(T,n'_p,n'_n)$:
\begin{equation}
\label{SE}
\Delta E_{j}^{SE}(n'_{p},n'_{n},T)= (A_j-Z_j) \Delta E_{n}^{SE}+ Z_j \Delta E_{p}^{SE} +\Delta E_{j}^{\rm eff.mass}\ . 
\end{equation}
For the potentials $\Sigma^0$ and $\Sigma$, we employ the  parametric formula
 for  RMF with the DD2 parameter set, Eqs.~(A1) and (A2) in \cite{typel10}.  
In the previous EOS \cite{furusawa17a}, the potentials are  
 evaluated consistently with  the RMF  with the TM1 parameter set employed for free nucleons.
The self-energy shifts for the light nuclei other than  $d, t, h$ and $\alpha$  are set to be zero. 
The parameters $a_{j/1}$, $a_{j/2}$, $a_{j/3}$, and $s_j$ are given in Table~I
in  \cite{typel10}.
The Coulomb energy shift is  obtained from the same calculation for Coulomb energy of heavy nuclei, Eq.~(\ref{eqclen}), by subtracting its vacuum limits  although the nuclear pasta phases are not considered  here.  
This mass evaluation for light clusters is stitched together 
by using a variety of different model assumptions, whereas mass fractions of light clusters in the FYSS EOS agree well with some experimental predictions \cite{hempel15,oertel17}.

\subsection{Translational energies of nuclei and thermodynamical quantities} \label{sectr}
The translational energy of nuclei is based on that for an ideal Boltzmann gas and is given by 
\begin{equation}%
\label{eq:tra}
 F_{i/j}^{t}= T \left\{ \log \left(\frac{n_{i/j}}{g_{i/j}(T) (M_{i/j} T/2\pi \hbar ^2 )^{3/2}   }\right)- 1 \right\} \left(1-\frac{n_B}{n_s}\right). 
\end{equation}
The last factor  takes account of the excluded-volume effect  in the same way as in  Lattimer and Swesty \cite{lattimer91}; each nucleus can move in 
the free space that is not occupied by baryons
and its translational motion is suppressed at high densities.
The internal degree of freedom $g_{i/j}(T)$ is defined as  
 $g_i(T)= (g_i^0 -1) \tau_i/{\rm sinh}\tau_i  +1$ for heavy nuclei and $g_j(T)= g_j^0$ for light nuclei
where  the spin degree of freedom of the ground state is defined by $g_{i/j}^0$. 
The factor $ \tau_i/{\rm sinh}\tau_i $ is incorporated washout of shell effects as in Eq.~(\ref{eq:shell}).

The abundances of nuclei as a function of $n_B$, $T$ and $Y_p$ are obtained by minimizing the model free energy in Eq.~(\ref{eq:total})
with 
respect to the number densities of nuclei and nucleons under baryon and charge conservations. 
The nuclear abundance of each nucleus is determined by 
 the nucleon chemical potentials  $\mu_{p/n}$ and the nuclear mass $M_{i/j}$  under chemical equilibrium.
Since both $\mu_{p/n}$ and  $M_{i/j}$   depend on the local proton and neutron densities $n'_{p/n}$ in our model, we solve  the equations relating $\mu_{p/n}$ and $n'_{p/n}$
 as well as the two conservation equations 
to determine the four variables: $\mu_p$, $\mu_n$, $n'_p$ and $n'_n$.

 Thermodynamical  quantities other than the free energy are derived by partial differentiations of the optimized free energy density, in which
all terms for in-medium effects 
are properly taken into account   to ensure thermodynamical consistency.
For example, the baryonic pressure  and the  entropy are obtained by the first derivative with respect to the baryonic density and the temperature, respectively.

\section{Results \label{sec:res}}
For comparison we present the results of  the new EOS based on the variational method   together with those of  the previous FYSS EOS based on the RMF with the TM1 parameter set \cite{furusawa17a}.
The free energy of free nucleons,  bulk energy of heavy nuclei and self-energy shifts of light nuclei are different between the two models.
The shell, surface and Coulomb energies of  heavy nuclei are also affected by the replacement of bulk properties such as the saturation densities of heavy nuclei. 
Table~\ref{tab1_bulk} summarizes the following bulk properties of nuclear matter at zero temperature: the nuclear saturation density of symmetric nuclear matter $n_{s0}$, internal energy per baryon  of symmetric matter at the saturation point $E_0$, compressibility $K$, symmetry energy $J$, 
and slope parameter in the density dependence of  symmetry energy $L$.
The variational EOS  is softer with smaller  values of $K$, $J$ and  $L$ \cite{togashi17}
although both EOS's can support the cold neutron stars
of masses
 $\sim$2~$M_{\odot}$ limit set by recent observations 
\cite{demorest,antoniadis13}.
On the other hand, 
radii of  1.4 $M_{\odot}$ neutron stars are quite different and $\sim$ 14.2 km  and 11.5 km,  for the RMF and variational EOS's, respectively \cite{togashi17}. 

\begin{figure}
\begin{center}
\includegraphics[width=11cm]{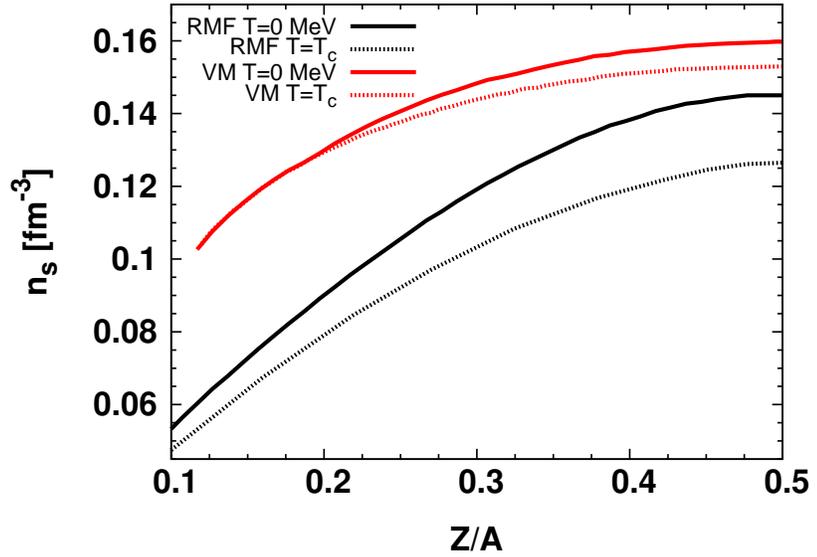}
\caption{Saturation  densities at zero temperature (solid lines) and those at  critical temperatures (dashed lines)  as a function of charge fraction 
for
the models based on the variational method  (red lines)
and  the  RMF (black lines)  }
\label{fig_sat}
\end{center}
\end{figure}

Figure~\ref{fig_sat} compares the saturation densities $n_{si}$ of heavy nuclei as a function of the charge fraction $Z_i/A_i$ 
at zero and critical temperatures, the latter of which  correspond to the temperatures
 where the saturation disappears.
The larger values of $J$ and $L$ in the  FYSS EOS lead to much lower saturation densities for  neutron-rich nuclei  compared with the  new EOS.
%
Not only the bulk energy but also the Coulomb and surface energies depend on $n_{si}$.

\begin{figure}
\begin{center}
\includegraphics[width=9cm]{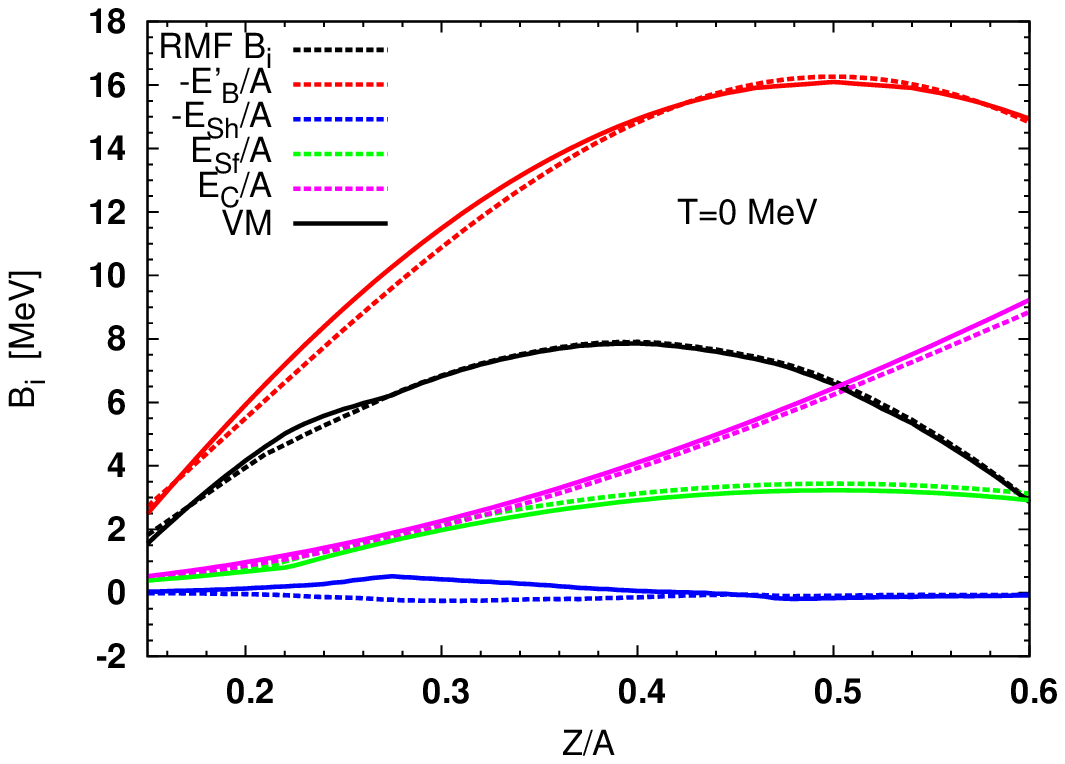}
\includegraphics[width=9cm]{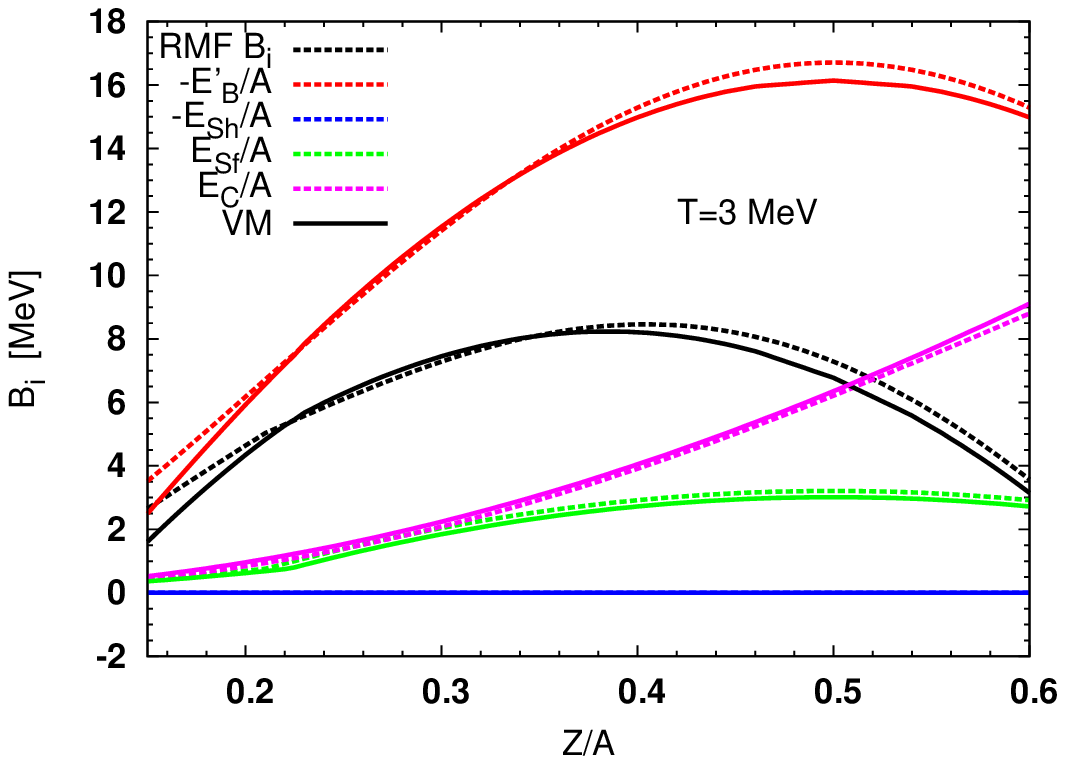}
\caption{Binding  (black lines), surface  (green lines), Coulomb (magenta lines), bulk (red lines) and shell  (blue lines) energies per baryon of the nuclei with $A_i=100$ 
 at the vanishing density and  $T$=0 MeV (left panel) and 3 MeV (right panel)  
 as a function of charge fraction 
for the new EOS   (solid lines)
and   the FYSS EOS (dashed lines).
Note that the sign of the bulk and shell energies is changed for convenience.
}
\label{fig_bind}
\end{center}
\end{figure}

Figure~\ref{fig_bind} displays the individual values of the bulk, shell, Coulomb,
 surface  and binding energies per baryon of nuclei with $A_i=100$ at two temperatures.
Note that the sign is changed in the former two and 
the baryon rest mass is subtracted from the bulk energy, which is defined as $E_i^{'B}$, and
 the binding energy per baryon is expressed as $B_i=- (E_i^{'B}+E_i^{C}+E_i^{Sf}+E_i^{Sh})/A_i$.
 The larger $n_{si}$ in the new EOS results in  smaller surface energies and larger Coulomb energies of heavy  nuclei.
Neutron-rich nuclei  in the FYSS EOS have smaller absolute values of the bulk energy due to the larger values  of  $J$ and $L$.
At the higher temperature ($T \gtrsim 3$~MeV), absolute values of the bulk energy become larger  due to the entropy contribution from nucleons in the nucleus, which  increases
 as the saturation density is reduced, especially for the  FYSS EOS.

\begin{figure}
\begin{center}
\includegraphics[width=7.0cm]{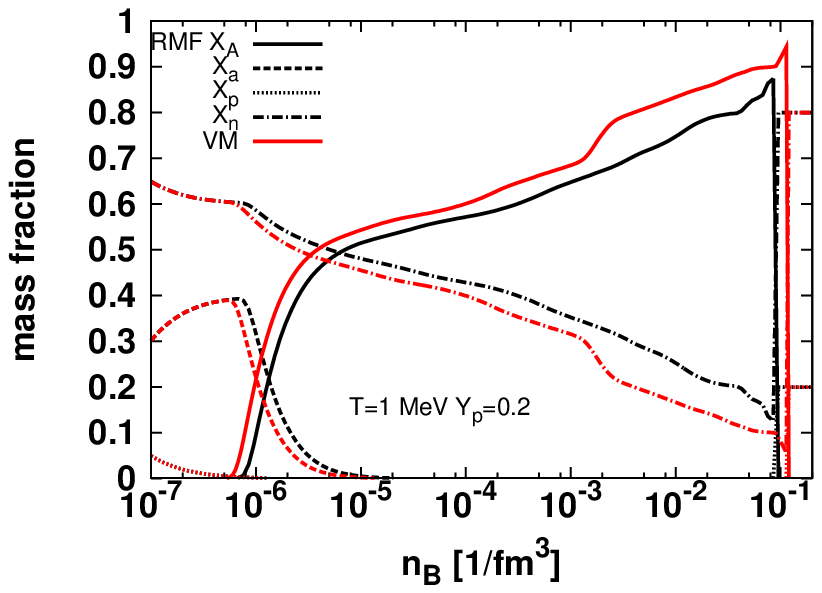}
\includegraphics[width=7.0cm]{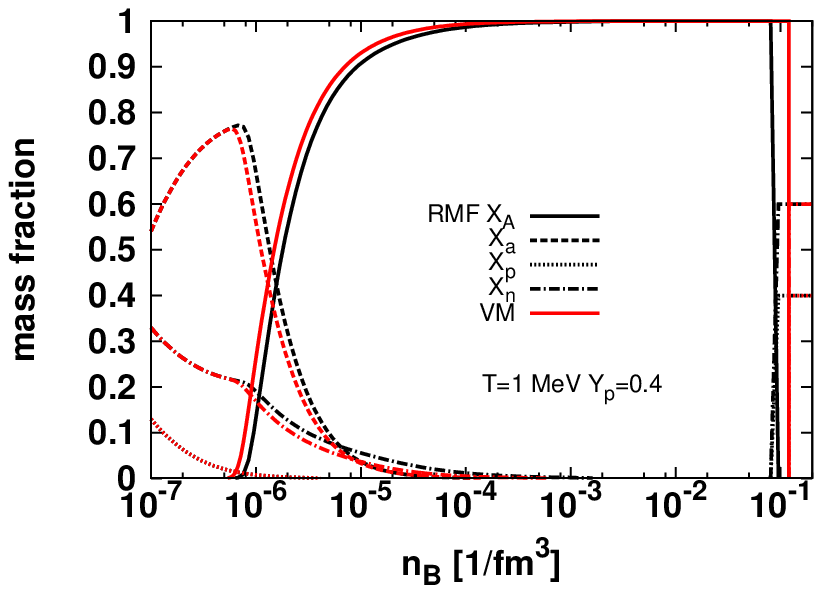}
\includegraphics[width=7.0cm]{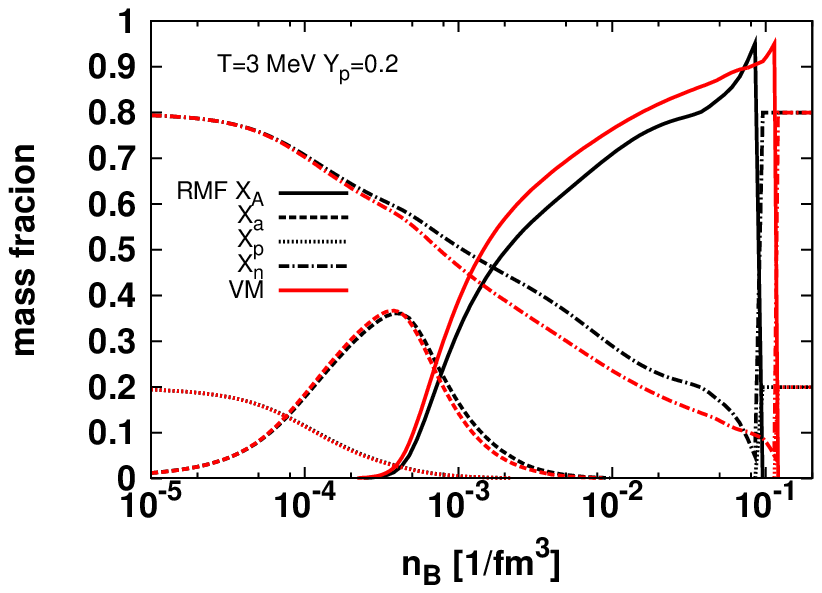}
\includegraphics[width=7.0cm]{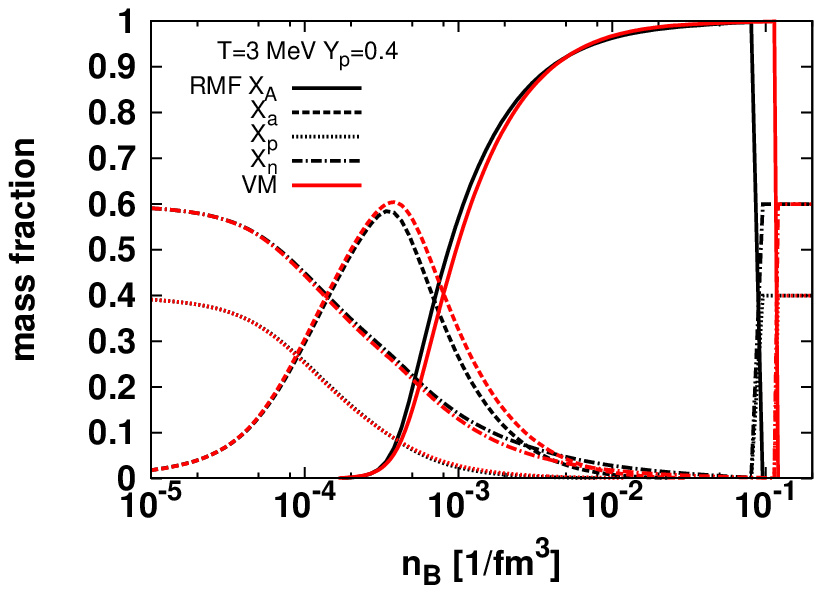}
\caption{Mass fractions of  heavy nuclei with $Z_i \geq 6$ (solid lines), $X_A$,  of light nuclei with $Z_i \leq 5$ (dashed lines), $X_a$, of free neutrons (dashed dotted lines), $X_n$,  and of free protons (dotted lines), $X_p$, 
for
the new EOS (red lines)
and    FYSS EOS (black lines)  
as a function of baryon number density,  $n_B$,
at  $T=1$ MeV (top row) and 3 MeV (bottom row)
and $Y_p=$ 0.2 (left column) and 0.4 (right column).
}
\label{fig_xhl}
\end{center}
\end{figure}

 Below we present the results   for $T$=~1, 3 and 5 MeV and  $Y_p=$ 0.2 and  0.4, typical values in the supernova core.
Figure~\ref{fig_xhl} shows  the mass fractions of free protons and neutrons, and of light and  heavy nuclei.
For the neutron-rich conditions of $Y_p=0.2$, the  mass fraction  of heavy nuclei in the new EOS is larger than that in the FYSS EOS.
This is because  neutron-rich nuclei have larger binding energies or smaller masses.
As for $Y_p=0.4$, the difference between the two EOS's is small compared with  the neuron-rich conditions, since the  bulk properties in the asymmetric nuclear matter 
are not much different between them.
We find that  
the mass fraction  of light nuclei decreases at some density and  the heavy nuclei  take their place at higher densities
 in both EOS's.
The mass fraction of free neutrons in the new EOS  tends to be
smaller than in the FYSS EOS due to the smaller
 value of $J$ in the former, which
  leads to large populations of neutron-rich nuclei. 
The mass fractions of free protons, on the other hand, are
 almost the same between the two EOS's.

\begin{figure}
\begin{center}
\includegraphics[width=12cm]{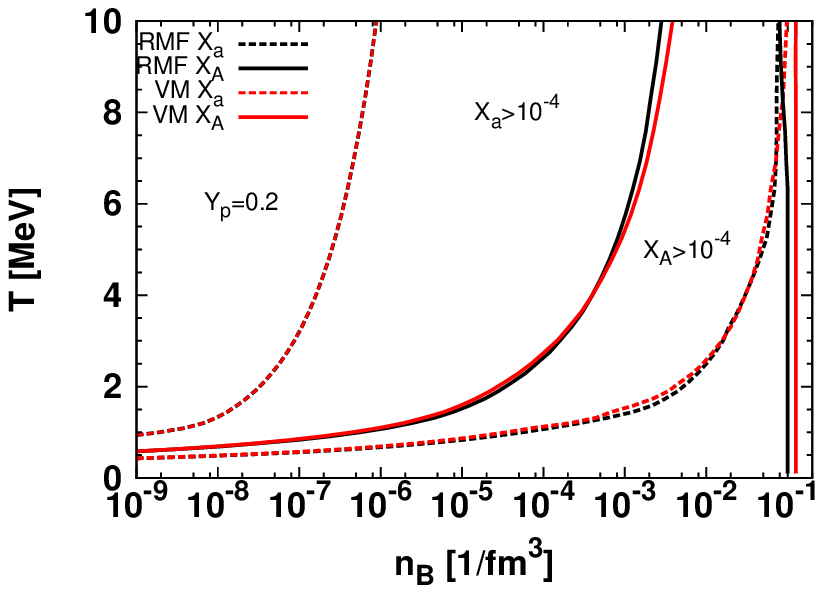}
\includegraphics[width=12cm]{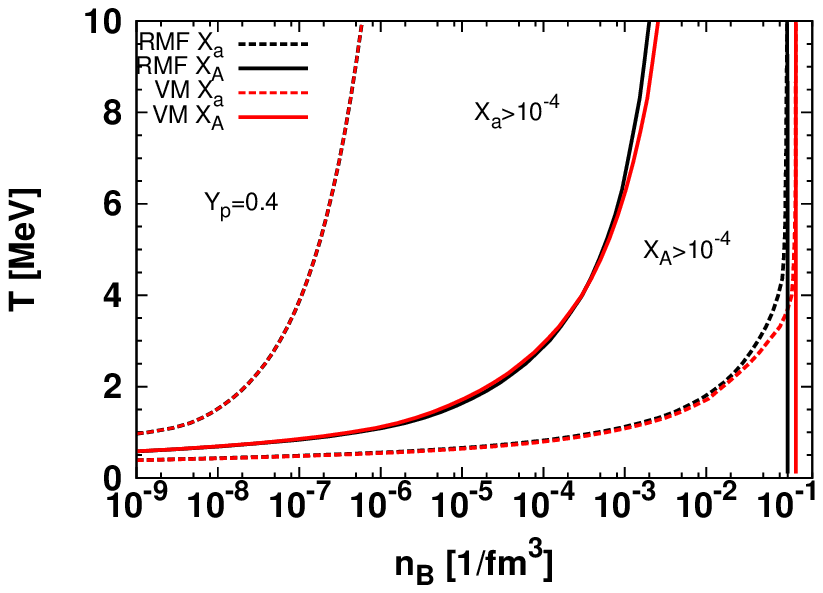}
\caption{Critical lines, on which  the  mass fractions
of light and heavy nuclei become  $X_a=10^{-4}$ and  $X_A=10^{-4}$,
for
the new EOS (red lines)
and   the  FYSS EOS (black lines)  
at $Y_p=0.2$ (top panel) and 0.4 (bottom panel). }
\label{fig_phase}
\end{center}
\end{figure}

Figure~\ref{fig_phase} displays the critical lines, where the mass fractions of heavy nuclei $(Z_i \geq 6)$, $X_A=\sum_{i} A_i n_i /n_B $ and of light nuclei $(Z_j  < 6)$,  $X_a=\sum_j A_j n_j$, 
 become $10^{-4}$.
At low temperatures  ($T \lesssim 3$~MeV), heavy nuclei appear at lower densities in the new EOS than in the  FYSS EOS whereas, at high temperatures, the order of the critical densities becomes opposite,  since the RMF we use yields  larger absolute  values of the  bulk energy,
 which originate  from the entropy contributions of nucleons in nuclei. 
On the other hand, both EOS's give  similar critical densities in Fig.~\ref{fig_phase}, at which  light nuclei show up. 
Note also the densities, at which nuclear pastas  melt into uniform nuclear matter depend on the saturation densities and those in the FYSS EOS are smaller.

\begin{figure}
\begin{center}
\includegraphics[width=7.0cm]{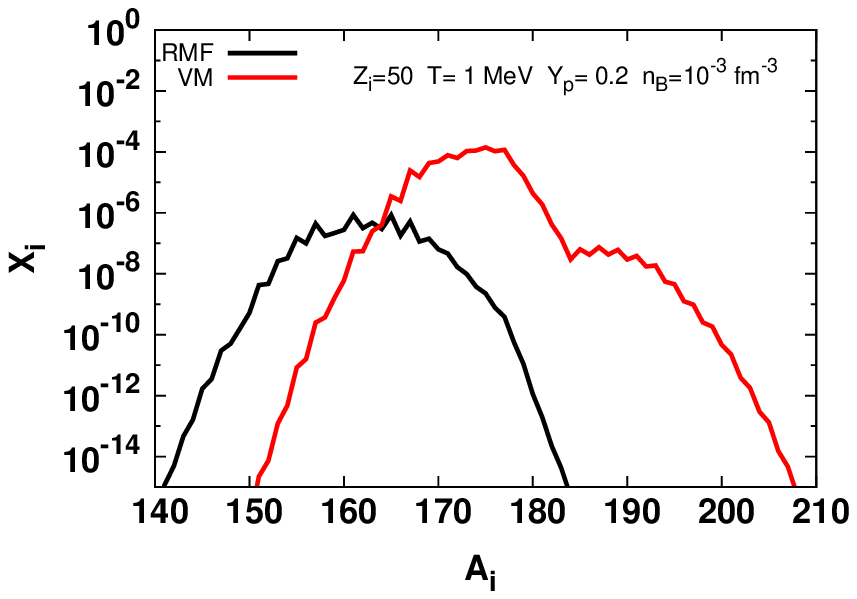}
\includegraphics[width=7.0cm]{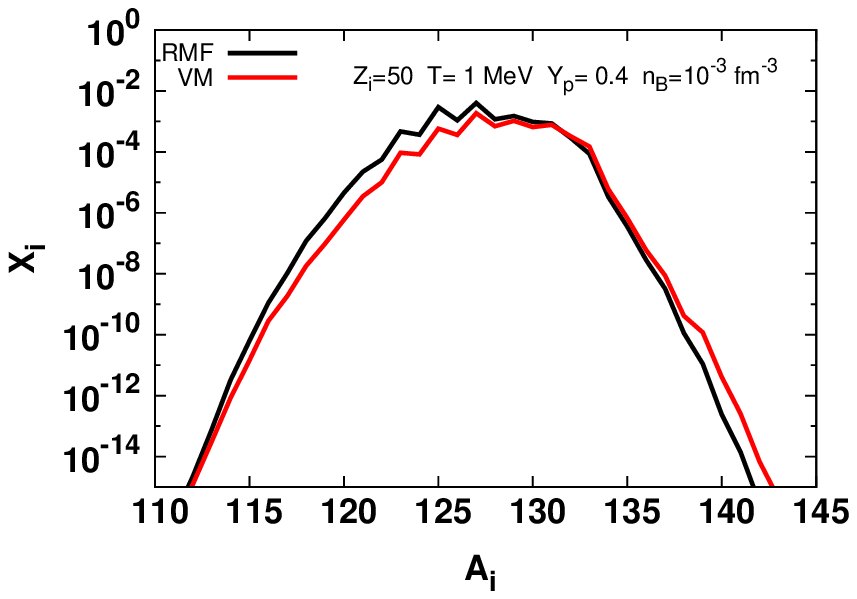}
\includegraphics[width=7.0cm]{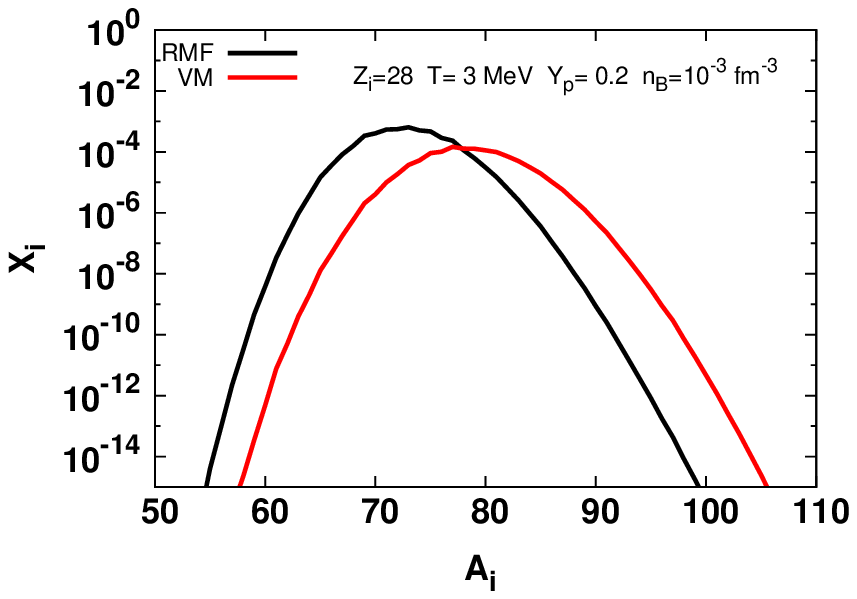}
\includegraphics[width=7.0cm]{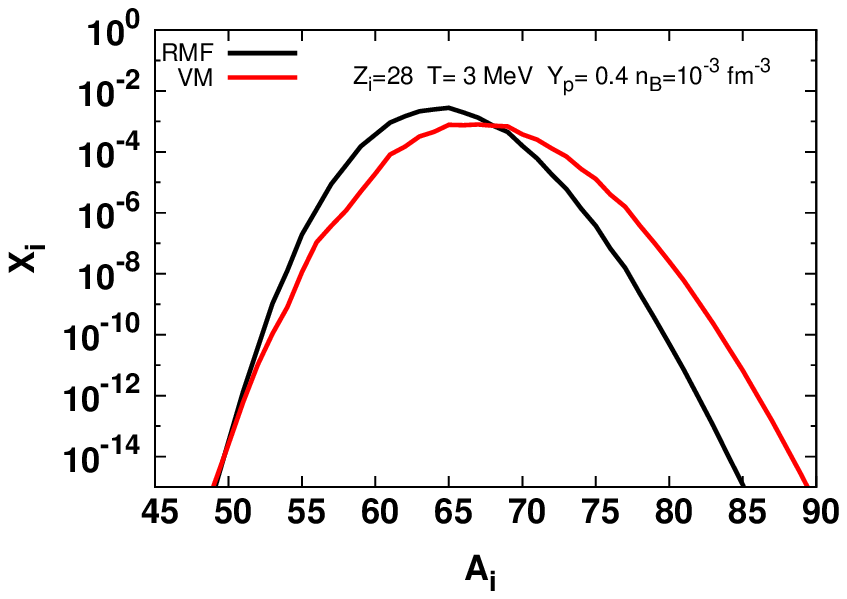}
\caption{Mass fractions of the  isotopes of Sn ($Z_i=50$) for  $T=1$ MeV (top row) and of Ni ($Z_i=28$) for $T= 3$ MeV (bottom row) as  a function of mass number 
for
the new EOS (red lines)
and the  FYSS EOS (black lines)  
at $Y_p=$ 0.2 (left column) and 0.4 (right column) and  $n_B=10^{-3}$~fm$^{-3}$.
}
\label{fig_iso}
\end{center}
\end{figure}

\begin{figure}
\begin{center}
\includegraphics[width=7.0cm]{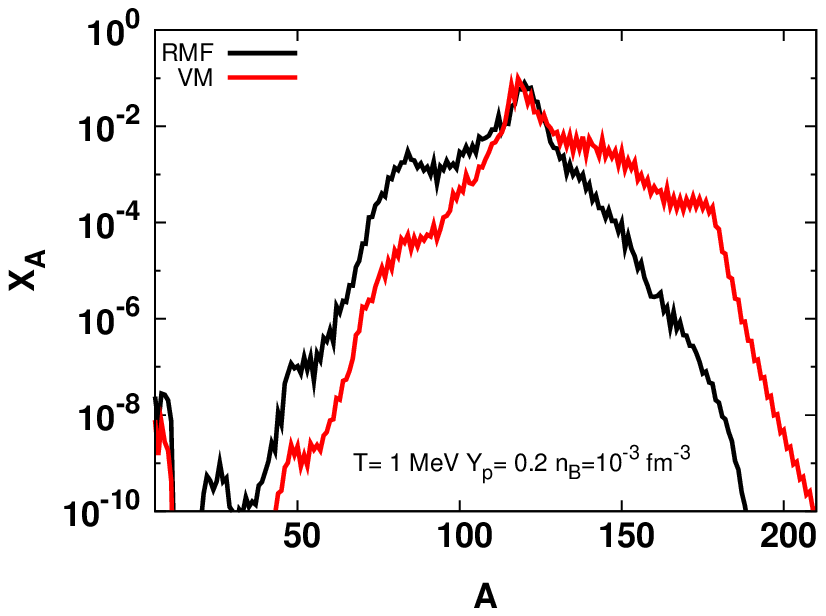}
\includegraphics[width=7.0cm]{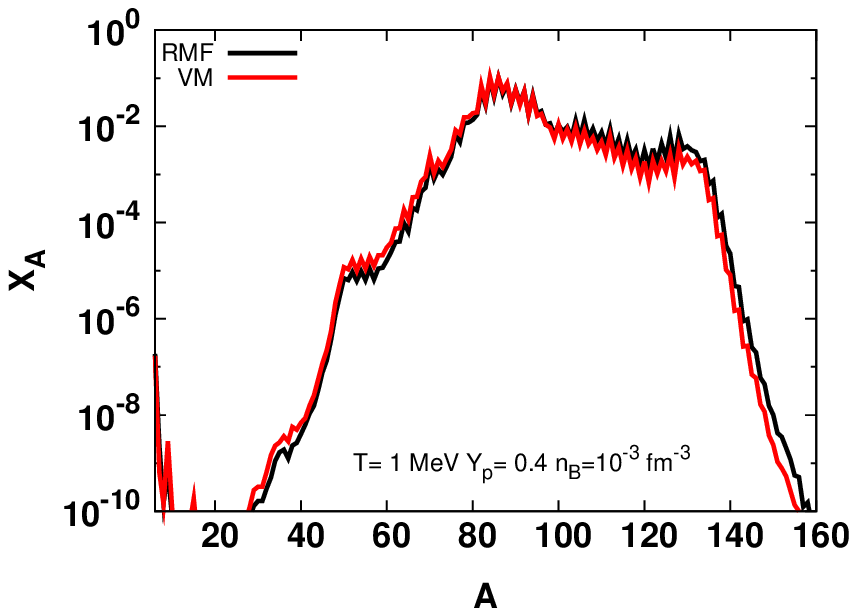}
\includegraphics[width=7.0cm]{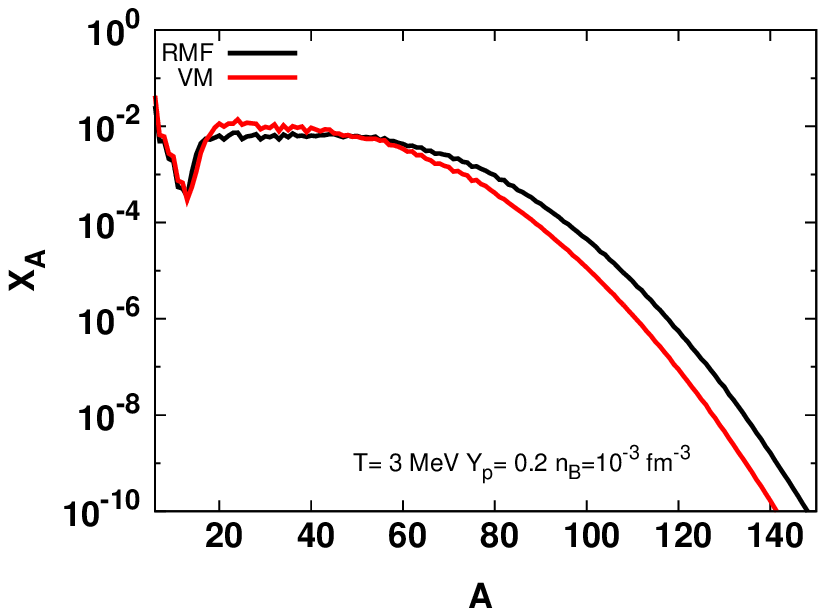}
\includegraphics[width=7.0cm]{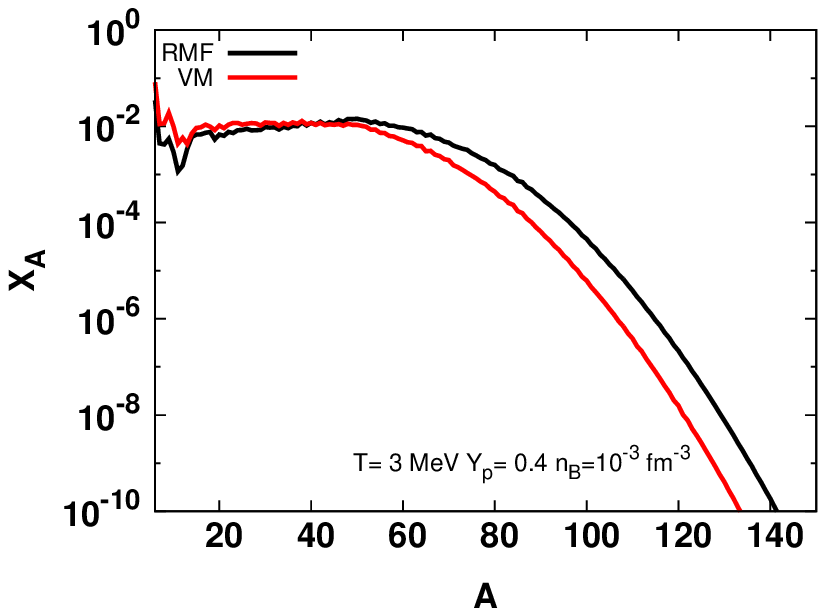}
\caption{Size distributions of  heavy nuclei as a function of mass number 
for
the new EOS (red lines)
and the   FYSS EOS (black lines)  
at  $T=1$ MeV (top row) and 3 MeV (bottom row),
$Y_p=$ 0.2 (left column) and 0.4 (right column) and  $n_B=10^{-3}$~fm$^{-3}$.
}
\label{fig_mad}
\end{center}
\end{figure}

 Figure~\ref{fig_iso}  shows the mass fractions of  the isotopes of Ni ($Z_i=28$) and Sn ($Z_i=50$).
In all cases considered there, larger amounts of 
 neutron-rich isotopes appear  in the new EOS than in the FYSS EOS 
due to the smaller symmetry energy.
Figure~\ref{fig_mad}, on the other hand,  displays the mass number distributions of heavy nuclei for four different conditions.
At  $n_B=10^{-3}$ fm$^{-3}$,  $T=1$ MeV and $Y_p=0.2$,  nuclei  with large mass numbers are more abundant in the new EOS,
since 
the smaller symmetry energy
allows  the nuclei in the vicinity of the neutron magic number $N=126$ to be more populated, which can also be confirmed in Fig.~\ref{fig_iso}.
For the other three conditions, the mass distribution  in the new EOS shifts slightly leftward
 compared with the FYSS EOS
owing to the  smaller surface  and larger Coulomb energies
 that in turn result from
the larger saturation densities in the former EOS.
The difference  between the two EOS's is particularly small at $T=1$ MeV and $Y_p=0.4$,
 which can be confirmed both in Figs.~\ref{fig_iso} and \ref{fig_mad}.
Note that zig-zag features remarkable
 at $T=1$ MeV, for which the shell effects are responsible,
 are almost washed out at $T=$ 3 MeV.  

\begin{figure}
\begin{center}
\includegraphics[width=7.0cm]{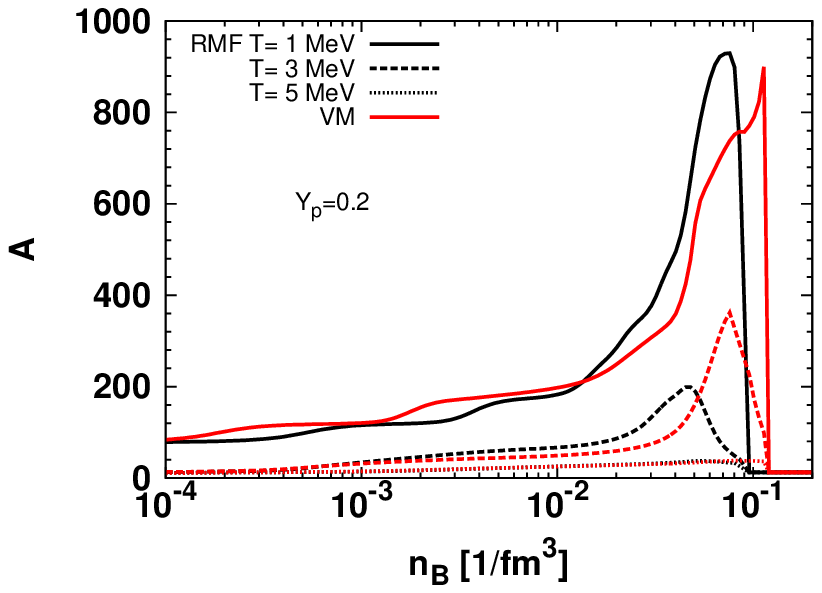}
\includegraphics[width=7.0cm]{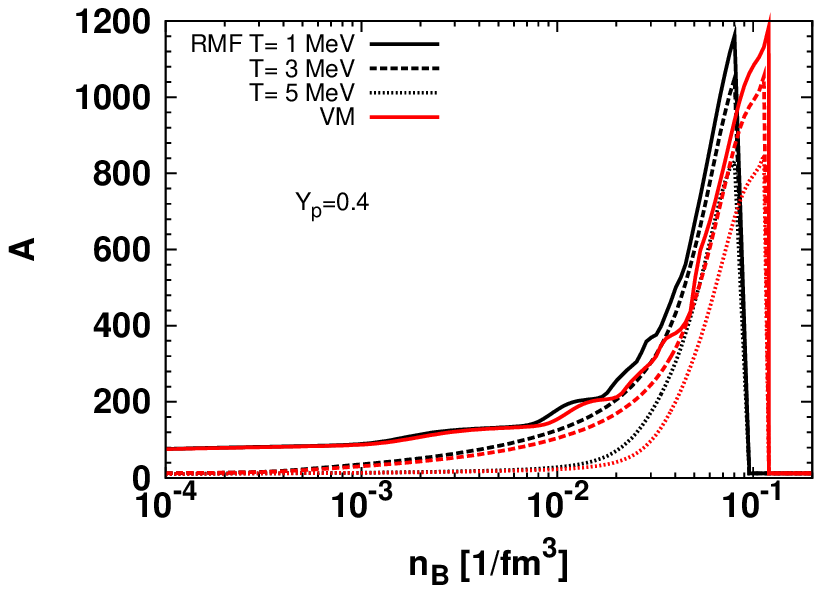}
\caption{
Average mass numbers of heavy nuclei with $Z \geq 6 $ 
as a function of  $n_B$
for
the new EOS (red lines)
and  the  FYSS EOS (black lines)  
at  $T=1$ MeV (solid lines), 3 MeV (dashed lines)  and 5 MeV (dotted lines)
for $Y_p=$ 0.2 (left panel) and 0.4 (right panel).
}
\label{fig_maa}
\end{center}
\end{figure}

Figure~\ref{fig_maa}  exhibits the average mass numbers of heavy nuclei as a function of density
 and  indicates that they are smaller in the new EOS than
  in the FYSS EOS in most cases due to the larger saturation densities  in the former. 
At $n_B\lesssim10^{-2}$ fm$^{-3}$, $T=1$ MeV and $Y_p=0.2$, the result is the other way round  because of  the larger abundance of neutron-rich magic nuclei for the new EOS.

\begin{figure}
\begin{center}
\includegraphics[width=7.0cm]{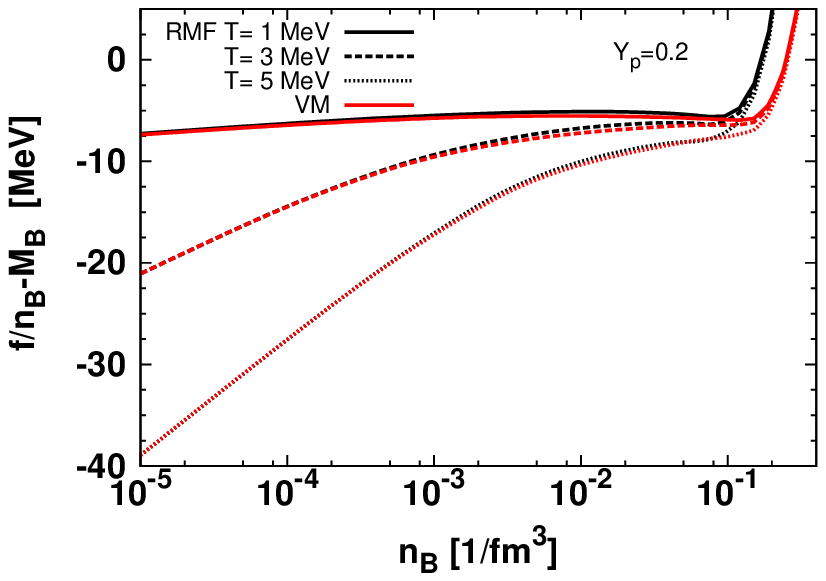}
\includegraphics[width=7.0cm]{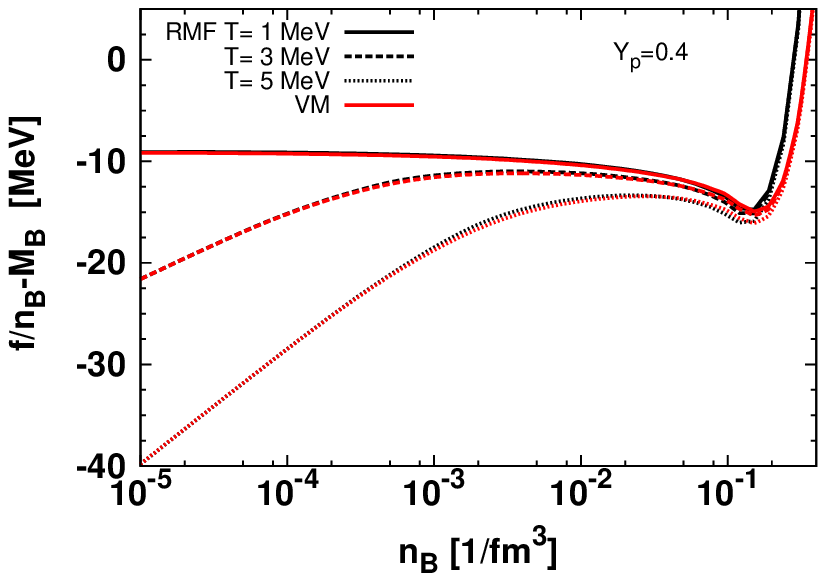}
\caption{Free energies per baryon relative to the baryon rest mass
as a function of  $n_B$
for
the new EOS (red lines)
and  the  FYSS EOS (black lines)  
at  $T=1$ MeV (solid lines), 3 MeV (dashed lines)  and 5 MeV (dotted lines)
for $Y_p=$ 0.2 (left panel) and 0.4 (right panel).
}
\label{fig_fre}
\end{center}
\end{figure}

\begin{figure}
\begin{center}
\includegraphics[width=7.0cm]{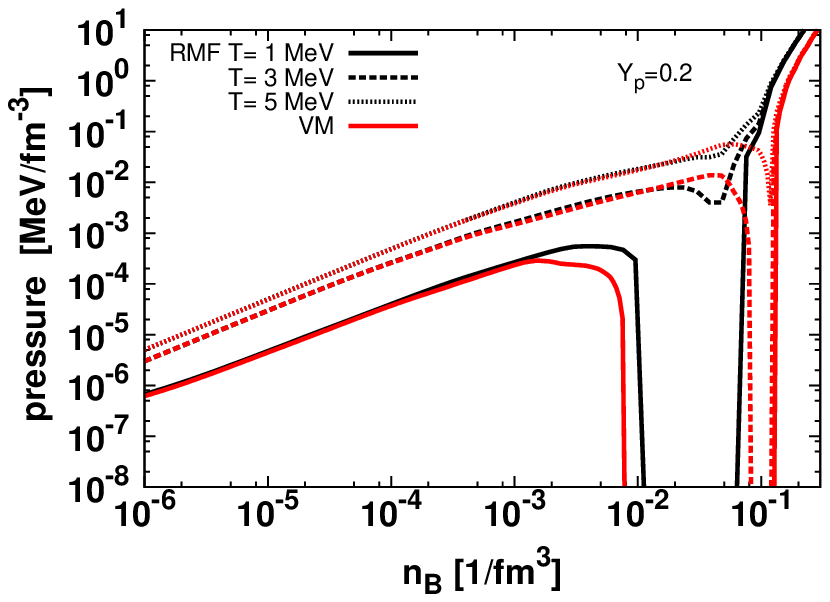}
\includegraphics[width=7.0cm]{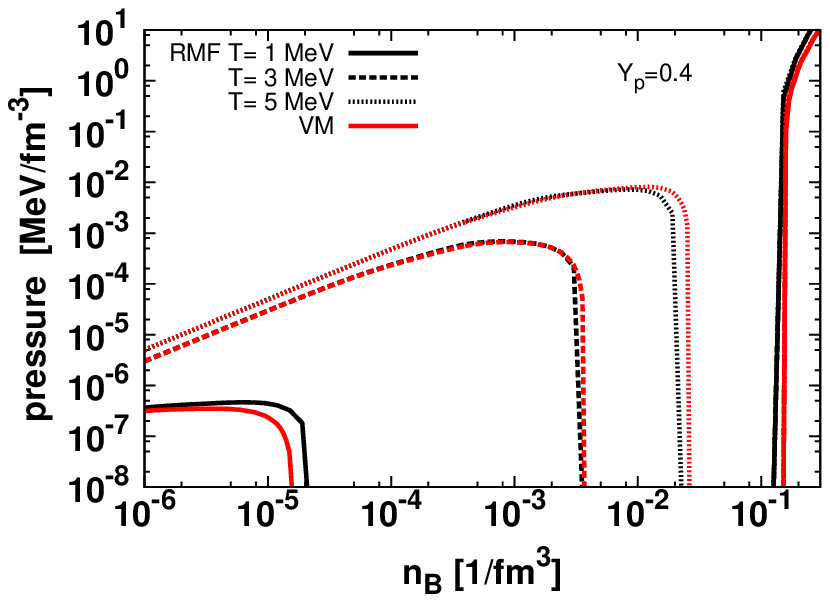}
\caption{  Baryonic pressures 
as a function of  $n_B$
for
the new EOS (red lines)
and the   FYSS EOS (black lines)  
at  $T=1$ MeV (solid lines), 3 MeV (dashed lines)  and 5 MeV (dotted lines)
for $Y_p=$ 0.2 (left panel) and 0.4 (right panel).
}
\label{fig_pre}
\end{center}
\end{figure}

\begin{figure}
\begin{center}
\includegraphics[width=7.0cm]{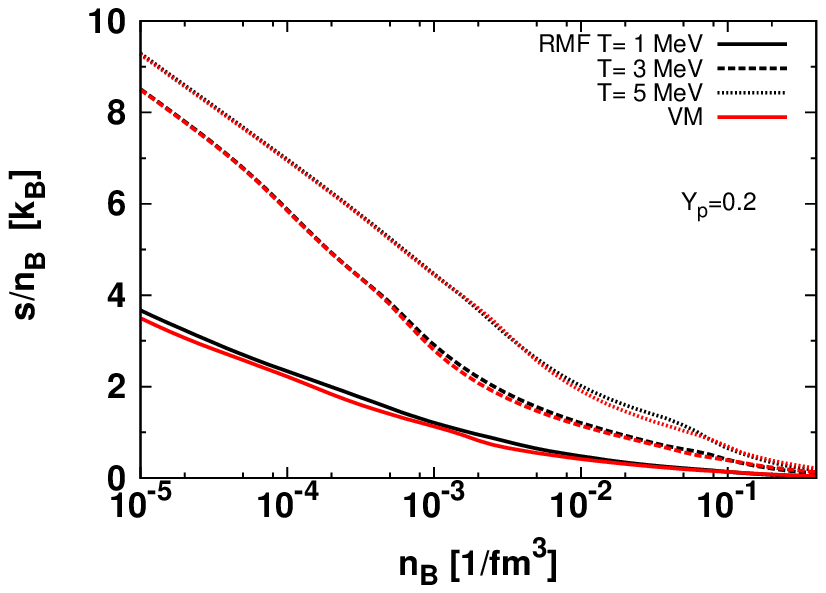}
\includegraphics[width=7.0cm]{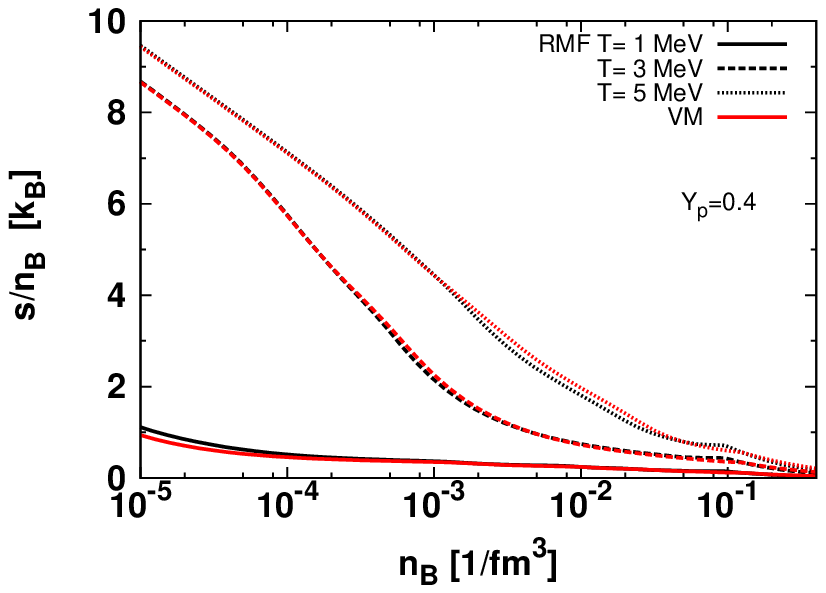}
\caption{ Entropies per baryon  
as a function of  $n_B$
for 
the new EOS (red lines)
and  the  FYSS EOS (black lines)  
at  $T=1$ MeV (solid lines), 3 MeV (dashed lines)  and 5 MeV (dotted lines)
for $Y_p=$ 0.2 (left panel) and 0.4 (right panel).
}
\label{fig_ent}
\end{center}
\end{figure}

Finally we compare thermodynamical quantities of relevance.
Figures~\ref{fig_fre}-\ref{fig_ent} present the free energy per baryon,  baryonic pressure
and entropy per baryon, respectively.
Around the saturation density, the FYSS EOS based on the RMF gives much larger free energies and baryon pressures for all  combinations of density, temperature and  proton fractions
considered there
 due to its stiffer bulk property. 
On the other hand, the difference of the free energies is not significant at low densities.
 At low  temperatures and high densities, the pressures become negative, since the Coulomb energies of heavy nuclei is decreased by the rise of electron density 
\cite{furusawa11}. 
The densities, at which the pressure becomes negative,
are influenced by the Coulomb energy and the nuclear composition and are hence a bit different between the two EOS's.
 Note that the total pressure of supernova matter is always positive owing to electron's contribution \cite{hempel10}.  
The entropy per baryon 
 is slightly affected by the difference in the mass fractions of nucleons and nuclei
as displayed in Fig.~\ref{fig_xhl}. 

\section{Applications to 1D  simulations of core-collapse supernovae \label{sec:simu}}
Implementing the new EOS in our neutrino radiation-hydrodynamics code, we have conducted spherically symmetric simulations of core collapse for a $11.2 \ M_{\odot}$ supernova progenitor \cite{woosley02}.
We have paid particular attention to the difference from the FYSS EOS \cite{furusawa17a}, with which we perform a comparable simulation with the same setting.
We designate  the simulation with the new EOS  as  the “VM” model
and the one with the FYSS EOS as the “RMF” model.

 Our code solves simultaneously the Boltzmann equations for neutrino transport, hydrodynamic equations for matter motions and Poisson equations for Newtonian gravity, 
 since they are all coupled with one another.
  We follow the dynamics up to $150 \ {\rm ms}$ after core bounce. Neutrino-matter interactions are 
treated in the same way as in  Nagakura et al.  \cite{nagakura16,nagakura17},
 which include updated sets of weak interactions and  of  electron capture (EC) rates for heavy nuclei  among other things \cite{juodagalvis10,langanke00,langanke03}.
They are  
calculated based on the nuclear populations obtained
from a nuclear statistical equilibrium EOS and with 
the previous version of the FYSS EOS \cite{furusawa13a} and
 tabulated as functions of $\rho$, $T$ and $Y_p$. 
 Note that
the same table is used for the simulations with the new and FYSS EOS's. It is true that the use of the EC rates based on the nuclear abundances derived with each EOS's is more consistent in these  simulations, but we are concerned with the difference not in the EC rates but in the EOS's themselves in this paper. We therefore decided to use the same EC table in this study. 
The results of fully consistent core collapse simulations  will be reported  elsewhere soon. For more technical details of our code, we refer readers to Nagakura et al. \cite{nagakura16}.

It turns out that 
the overall dynamics are not much different between the two simulations. 
In particular, the shock radius is surprisingly similar to each other as a function of time
even though the distributions of $\rho$, $T$ and $Y_p$ are not so identical as described in detail below. 
This seems to be an accident, with different effects canceling each other. 
The use of the same EC rates may also be contributing to this similarity. 
In neither case, we obtain an explosion as expected for spherical models. If one looks into details, 
however, some  differences become apparent, to which we shift our attention hereafter.

\begin{figure}
\begin{center}
\includegraphics[width=7.0cm]{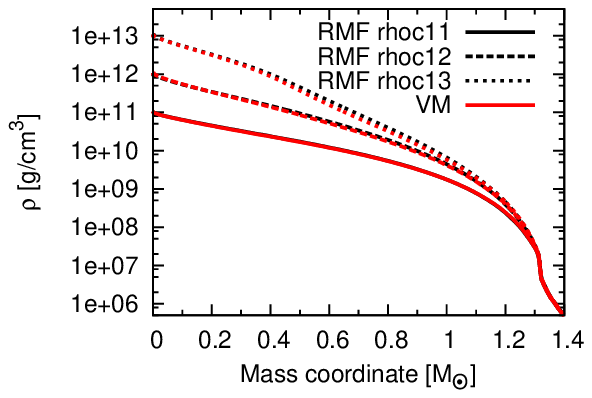}
\includegraphics[width=7.0cm]{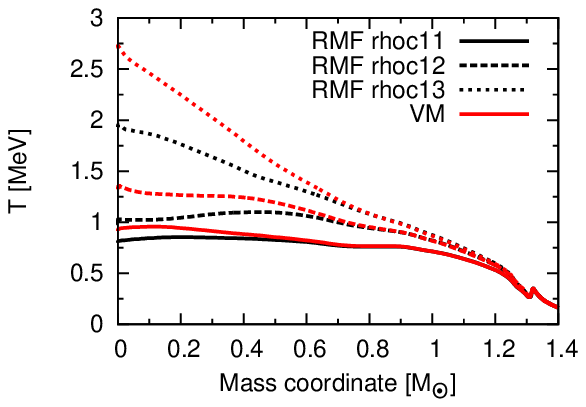} \\
\includegraphics[width=7.0cm]{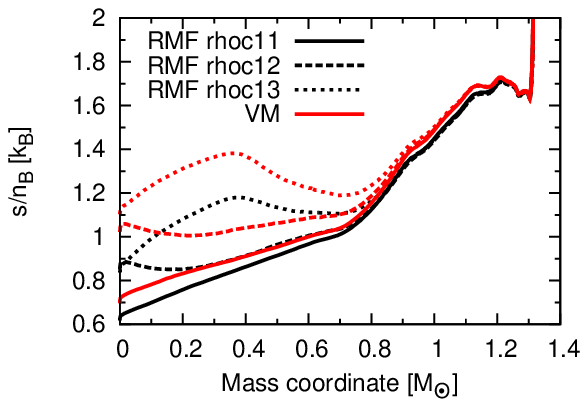}
\includegraphics[width=7.0cm]{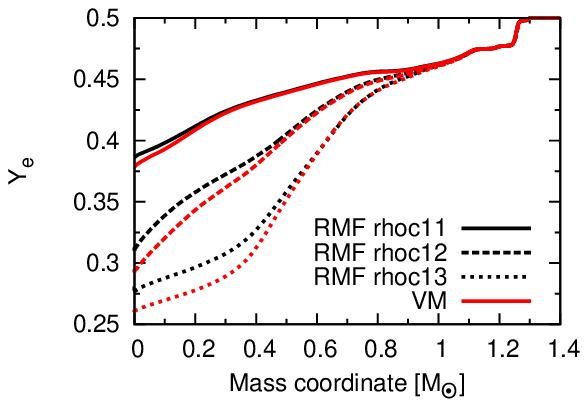}  \\
\includegraphics[width=7.0cm]{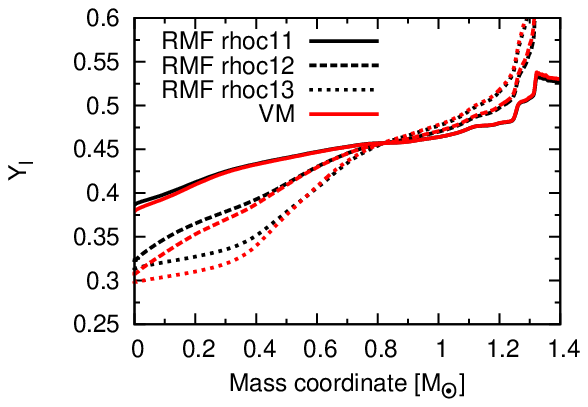}
\includegraphics[width=7.0cm]{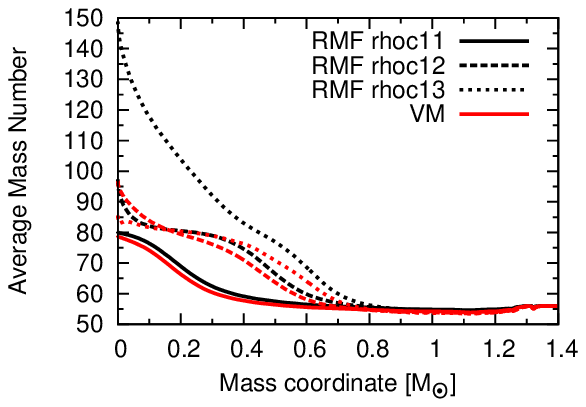}
\includegraphics[width=7.0cm]{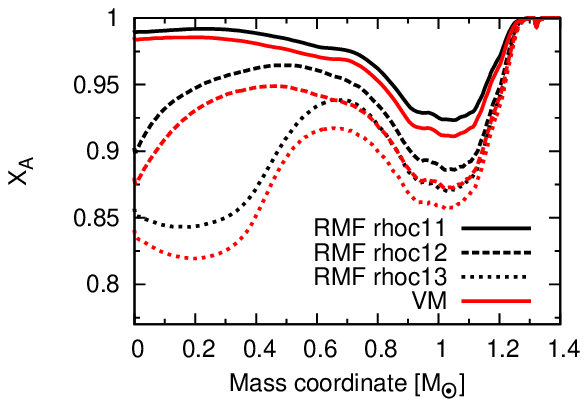}
\includegraphics[width=7.0cm]{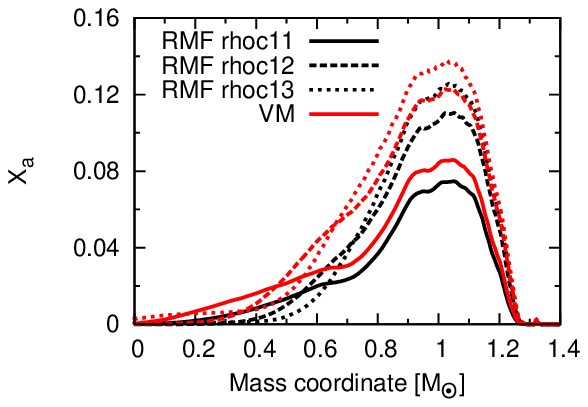}
\caption{Radial distributions of several  quantities of relevance obtained by 1D spherically symmetric simulations for the VM  (red lines)
and   RMF  (black lines) models at different times in the pre-bounce phase when the central density reaches $10^{11}$ (solid lines), $10^{12}$ (dashed lines) and $10^{13}$ (dotted lines) ${\rm g/cm}^3$.
The horizontal axis is the mass coordinate.
}
\label{fig_prebounce}
\end{center}
\end{figure}

 In Fig.~\ref{fig_prebounce}, we display radial profiles of several important quantities as a function of mass coordinate at three different times in the collapse phase, i.e.,  when the central density ($\rho_{c}$) reaches $10^{11}, 10^{12}$ and $10^{13} \ {\rm g/cm}^3$. In these figures, we can immediately recognize the differences in the temperature and entropy profiles in the central region between the two models, which are particularly significant after the central density reaches $10^{12} \ {\rm g/cm}^3$. 
This higher entropy for the new EOS  originates from
 smaller mass fractions of heavy nuclei  in the  core matter with $Y_p \sim 0.46$ at the beginning of the collapse.
The reason is as follows: in
the matter with heavy nuclei and $\alpha$ particles dominating over free nucleons,
 the mass fraction of heavy nuclei  is approximately given as $X_A \sim (0.5-Y_p)/(0.5-\bar{Z}/\bar{A})$, 
 which is  in turn obtained from the baryon number conservation  $ X_A + X_{\alpha} \sim 1$ and the charge neutrality $ (\bar{Z}/\bar{A}) X_{A} + 0.5 X_{\alpha} \sim Y_e=Y_p$, where  $\bar{Z}$ and $\bar{A}$ are the average atomic and mass numbers  of heavy nuclei.
 For a given $Y_p$,
  neutron-rich nuclei are more abundant, 
implying a smaller $\bar{Z}/\bar{A}$, in the new EOS,
  because of the smaller symmetry energy as discussed in the previous section.
This is borne out in the bottom left panel of Fig. \ref{fig_prebounce}. The matter 
 around the mass coordinate of  $1.0  \ M_{\odot}$
is similar to the matter at the center at the initial time. It is evident that the  
 VM model gives a
  smaller $X_A$ and a larger $X_a \sim X_{\alpha}$, and hence a higher entropy per baryon 
although $\rho$, $T$ and $Y_p$ are almost identical between the two models.
In the early phase of core collapse ($\rho_c \sim  10^{10} \ \rm{g/cm^3}$),
the temperature of the central region increases more rapidly by  adiabatic compression in the VM model with the  higher entropy than  in the RMF model. 
The higher temperature accelerates electron captures and deleptonizes the core more quickly (see the radial profile of $Y_e$ in Fig.~\ref{fig_prebounce}). The smaller electron fraction decreases the degeneracy pressure of electrons, 
accelerating the collapse, raising   the temperature further, and resulting in rather large deviations in temperature and entropy (see the corresponding panels). Note that the entropy is changed by electron captures and accompanied emissions of neutrinos. 

 It is also important to note that such  high temperatures and small electron fractions reduce the mass fraction of heavy nuclei
further, enhancing the mass fractions of  nucleons and light nuclei instead. 
The matter in the central region ($\lesssim  0.6 \ M_{\odot}$) 
consists of  heavy nuclei with $\bar{A} \gtrsim 60$, light nuclei  and dripped neutrons
when $\rho_c$ exceeds $\sim$10$^{12}$ g/cm$^3$. For such matter with $T\lesssim$ 3 MeV and $Y_p(=Y_e) \lesssim 0.4$,
the new EOS, in contrast to the previous case, tends to give 
 greater values to $X_A$ than the FYSS EOS,
 as shown in the previous section. 
Owing to the higher $T$ and lower $Y_p$ ($=Y_e$) in the VM model  than  in the RMF model, however,
the order is reserved, keeping $X_A$ lower 
in the dynamical simulations of the VM model.
The smaller mass fractions  of heavy nuclei imply lower opacities from neutrino-nucleus coherent scatterings, the  dominant  source of opacity during collapse, meaning that the deleptonization is 
further accelerated in the
 VM model. 
This in turn  makes the mass of  the inner core smaller. As is well known, the small mass of the inner core 
is disadvantageous  for the outward propagation of  the shock wave  generated by core bounce, since the gravitational energy liberated and hence the initial shock energy tend to be smaller,
 and the outer core, the obstacle for shock propagation, becomes larger \cite{yahil82,bruenn85}.

\begin{figure}
\begin{center}
\includegraphics[width=7.0cm]{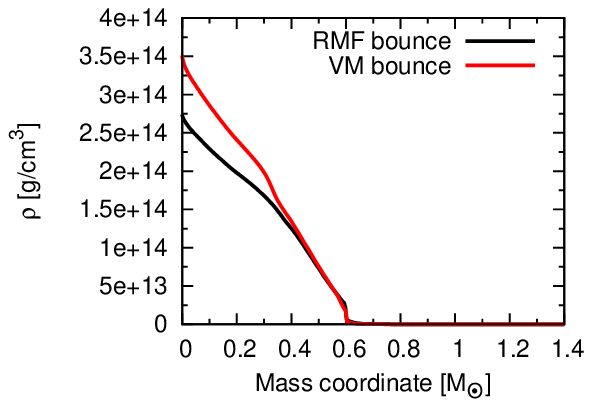}
\includegraphics[width=7.0cm]{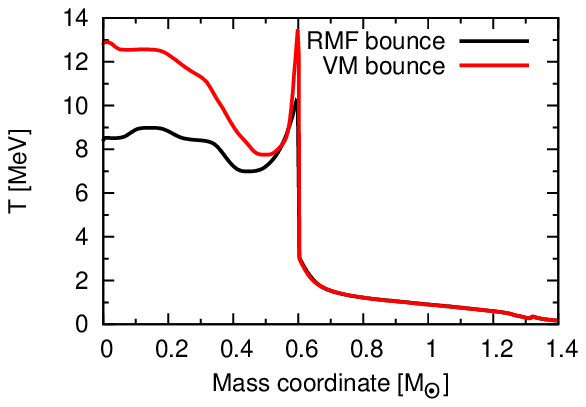} \\
\includegraphics[width=7.0cm]{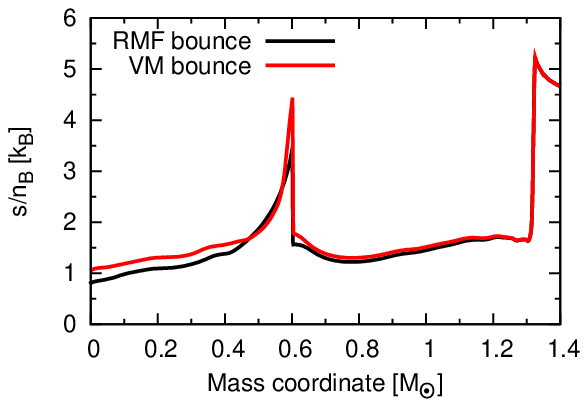}
\includegraphics[width=7.0cm]{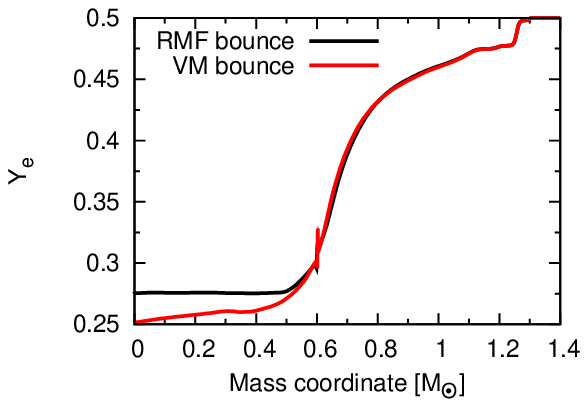} \\
\includegraphics[width=7.0cm]{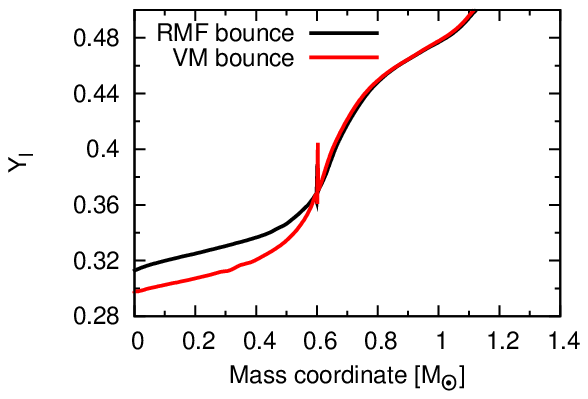}
\includegraphics[width=7.0cm]{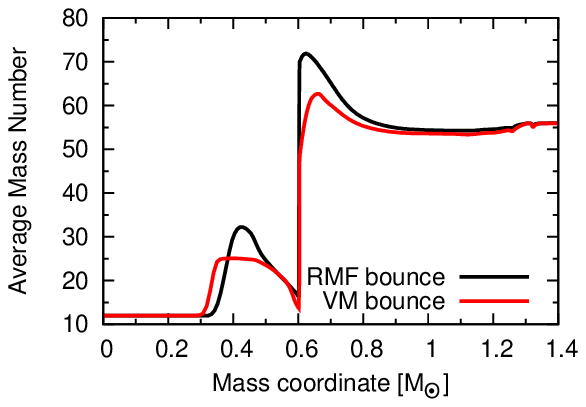}  \\
\includegraphics[width=7.0cm]{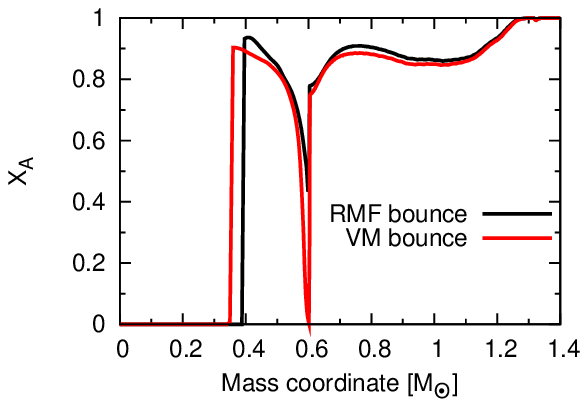}   
\includegraphics[width=7.0cm]{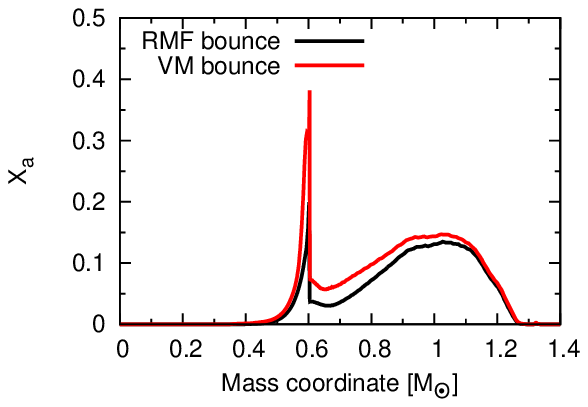} 
\caption{Same as Fig.~\ref{fig_prebounce} but at core bounce.  
}
\label{fig_bounce}
\end{center}
\end{figure}

\begin{figure}
\begin{center}
\includegraphics[width=7.0cm]{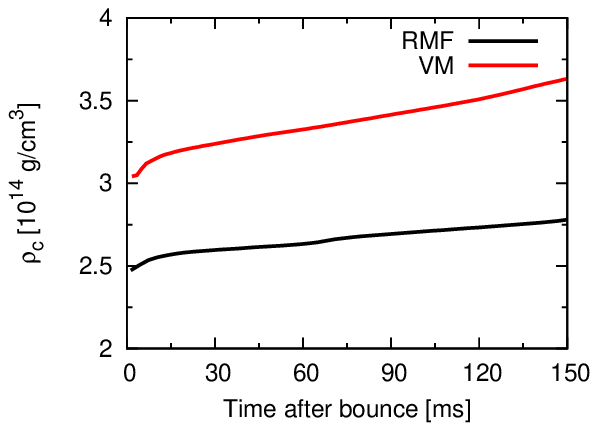}
\includegraphics[width=7.0cm]{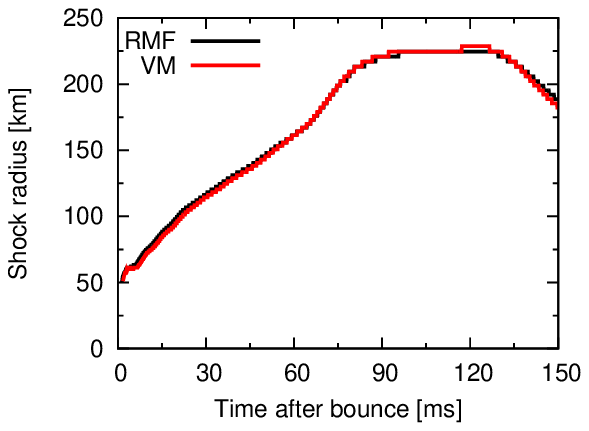}
\caption{Time trajectories of the central density (left panel) and shock radius (right panel) for both models.
}
\label{fig_postbounce_rhoandshock}
\end{center}
\end{figure}

Figure~\ref{fig_bounce} shows the radial profiles of the same quantities as in Fig.~\ref{fig_prebounce} but at the time of core bounce,
 which we define in this study as the time when the shock wave reaches the mass coordinate  of  $0.6 \ M_{\odot}$. As discussed in the previous paragraph, the stronger deleptonization for the  VM model gives  the central electron fraction $\sim 0.25$ at bounce, which is $\sim 10\%$ smaller  than that for the RMF model. We find, however, that 
the shock wave is  stronger in the VM model  than in the RMF, which can be understood 
from the greater jump of entropy at  the shock wave for the VM model. This result 
may seem at odds with our previous statement that 
 the shock energy tends to be lower for the smaller inner core.
This is true only when the EOS is identical, though. 
In fact, the new EOS is softer than the FYSS EOS as mentioned earlier.
Then the inner core becomes not only less massive but also more compact, 
which is corroborated in the top left panel of the figure. The central density at core bounce is indeed higher in the VM model. The smaller radius of the inner core means a larger liberation of  gravitational energy.  Indeed, 
this is sufficient to compensate for the smaller mass of inner core and leads to  
the similar trajectory of shock wave at later times in the two models
 (see the right panel in Fig.~\ref{fig_postbounce_rhoandshock}).

\begin{figure}
\begin{center}
\includegraphics[width=7.0cm]{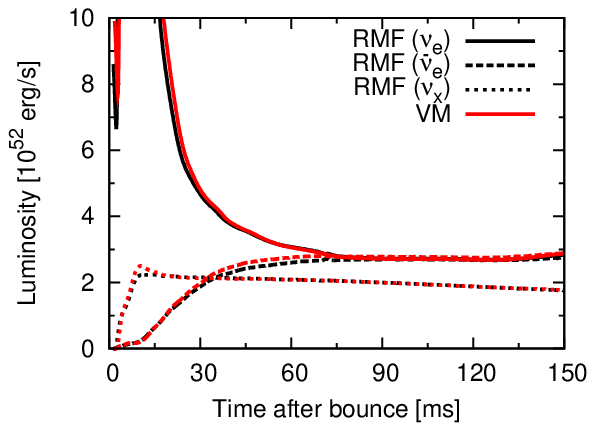}
\includegraphics[width=7.0cm]{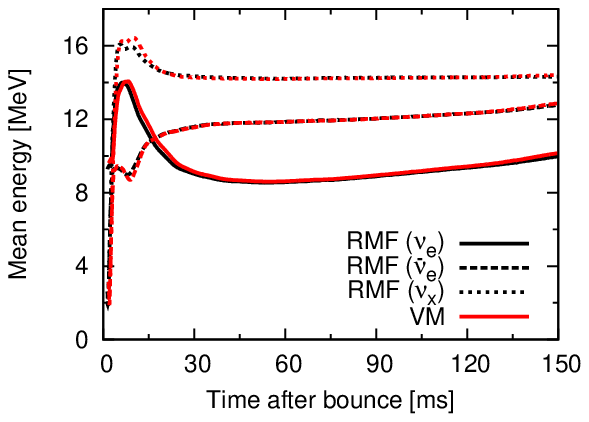}
\caption{Time evolutions of the luminosities  (left panel) and mean energies (right panel) of neutrinos. Note that the mean energy is  defined as the ratio of  the energy density to the number density of neutrinos.
}
\label{fig_postbounce_neutrinos}
\end{center}
\end{figure}

The left panel in Fig.~\ref{fig_postbounce_rhoandshock} compares the time evolutions of the central density in the post-bounce phase between the two models. The new EOS has
consistently higher values  than the RMF model, an indication that the former is softer and the inner core is more compact.
In spite of the rather large difference in the central density
between the two models, the time evolutions of the  shock radii are quite similar to each other (see the right panel in the same figure). It should be noted, however, that this  coincidence might be 
an artifact of our use of 
 the same EC rates for heavy nuclei, which
is certainly inconsistent in the VM  and RMF models, 
not correctly reflecting the differences in the nuclear composition.
In fact, the outer cores have very similar deleptonization  histories during the collapsing phase between the two models (see the panel for $Y_e$ in Fig.~\ref{fig_prebounce}), which means that 
  the shock waves propagate through very similar outer cores.
It is also important to note that the luminosity and mean energy of neutrinos are also almost identical between the two models (see  Fig.~\ref{fig_postbounce_neutrinos}), meaning that neutrino coolings  are almost the same, which also contributes to the similar  evolutions of shock radius. 

\begin{figure}
\begin{center}
\includegraphics[width=7.0cm]{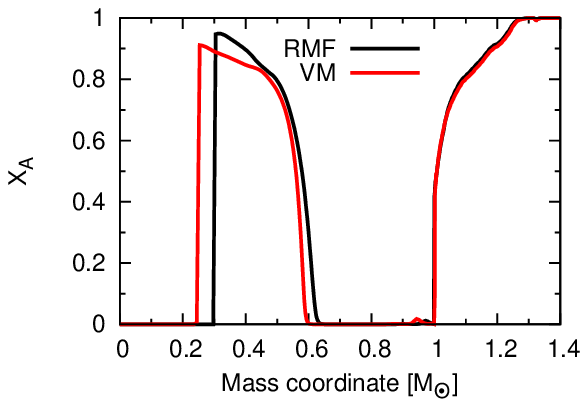}
\includegraphics[width=7.0cm]{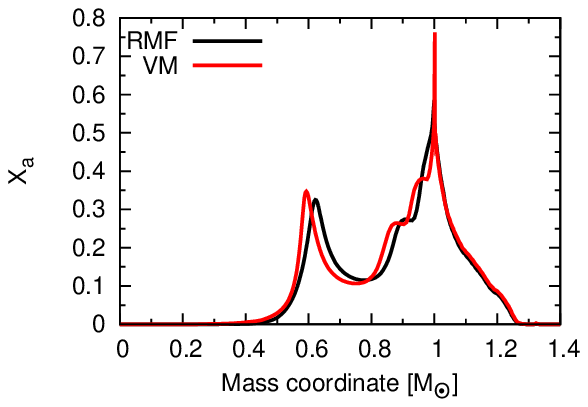}
\caption{Radial distributions of the mass fraction of heavy  (left) and light (right) nuclei at the time when the shock reaches $1.0 \ \rm{M}_{\odot}$ on the mass coordinate for the VM and RMF models.
}
\label{fig_postbounce_Msh1}
\end{center}
\end{figure}

\begin{figure}
\begin{center}
\includegraphics[width=7.0cm]{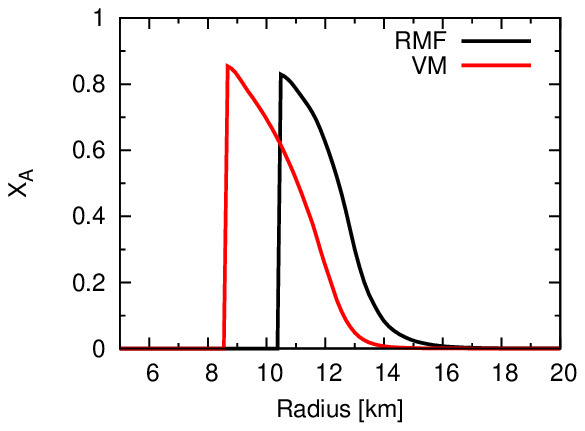}
\includegraphics[width=7.0cm]{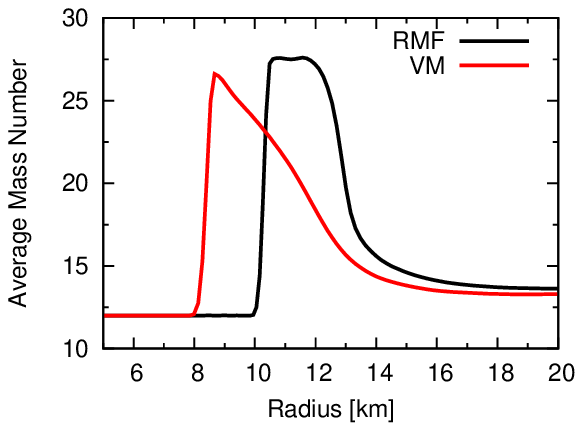}
\includegraphics[width=7.0cm]{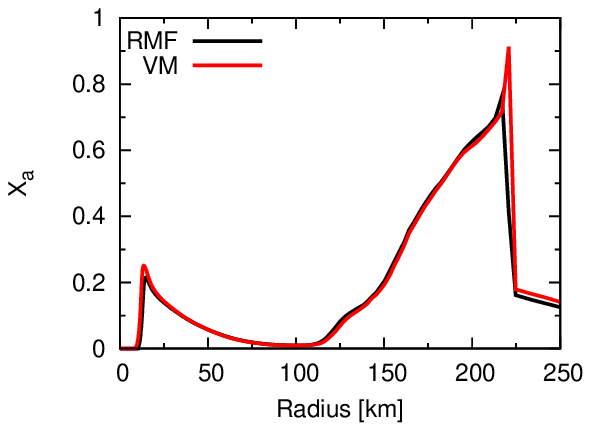}
\caption{
Radial distributions of the mass fraction of heavy (top left)  and light (bottom) nuclei
and the
 average mass numbers
(top right) for the VM (red lines) and RMF (black lines) models at 100 ms after core bounce. Only the vicinity of the PNS surface (5-20 km) is shown in the top panels whereas  the entire post-shock region is exhibited in the bottom panel.
}
\label{fig_postbounce_tp100ms}
\end{center}
\end{figure}

It is interesting to note that substantial amounts of heavy nuclei exist around the PNS 
(see the left panel in Fig.~\ref{fig_postbounce_Msh1}) and they survive for more than $100$ ms after  bounce (see the upper left panel in Fig.~\ref{fig_postbounce_tp100ms}). We expect that they might affect the diffusion of neutrinos,
 since the coherent scattering by heavy nuclei 
gives far greater opacities than scatterings on free nucleons or light nuclei.
Incidentally,  
 there is a non-negligible mass fraction of light nuclei just behind the shock wave
(see the right panel of Fig.~\ref{fig_postbounce_Msh1}). This may indicate  that weak interactions between neutrinos and these  light nuclei should not be ignored in the post-shock region and may play an important role in the timing of the neutrino burst. 
We will address this issue by  incorporating the relevant weak interactions with light nuclei in simulations \cite{furusawa13b,nasu15,fischer16}.
It should be noted that non-negligible  mass fractions of light nuclei appear just outside  the PNS. In fact, their mass fraction sometimes overwhelms that of nucleons. 
These light nuclei, both behind the shock wave and near the PNS, may hence have the potential to change the neutrino cooling/heating  (see also \cite{furusawa13b}) and affect  shock revival eventually. The detailed study on these effects will be  presented in our forthcoming paper.

\section{Summary and discussion \label{sec:conc}}
We have constructed a new equation of state for use in supernova simulations. It  includes the full ensemble of nuclei
and is derived from the EOS for uniform nuclear matter that employed realistic nuclear forces in the variational method.
This should be contrasted with standard EOS's, which include
a single representative nucleus and treat the nuclear interactions 
phenomenologically with either 
 Skyrme type interactions or 
meson-exchange models.
In the calculation of  nuclear masses, 
we  have taken into consideration
various in-medium effects such as  the smearing of shell effects in heavy nuclei
 and the self- and Pauli-energy shifts in light nuclei  at high densities and/or temperatures.

For some representative conditions in supernovae, we have compared the new EOS and our previous one: the FYSS EOS, in which the RMF EOS with the TM1 parameter set is employed. 
At  $T =1$ MeV and  $Y_p=0.4$,  typical values at the center 
in the initial phase of   collapse, the difference between the nuclear compositions in the two EOS's is not so large,
 since normal nuclear statistical equilibrium is expected to prevail with negligible in-medium effects. 
For  more neutron-rich conditions  ($Y_p\sim0.2$),  
 the smaller symmetry energy in the new  EOS provides larger mass fractions of heavy nuclei than the FYSS EOS.
At higher temperatures ($T \gtrsim 3$ MeV), 
the larger  entropy contribution 
 leads to
  larger total  mass fraction of heavy nuclei  in the FYSS EOS than in the new EOS.
It is also found that  the new EOS tends to have larger mass fractions of nuclei with small mass numbers and charge fractions owing to the larger saturation densities and the smaller symmetry energy.
The abundance of light nuclei, on the other hand, 
is not much changed by the replacement of the previous EOS with the new one
although it is somewhat affected by the difference in the mass fractions of  heavy nuclei. This is because in the new EOS the  self- and Pauli-energy shifts are not evaluated consistently with the EOS for uniform nuclear matter.

Employing the two EOS's, we have performed dynamical simulations of  core-collapse
under the assumption of spherical symmetry. 
We have found that  the new EOS gives  smaller mass fractions  of heavy nuclei and higher entropies
for conditions typical in the early phase of core-collapse. This is due to the lower symmetry energy in the new EOS, which leads to slightly lower values of the charge fraction. Owing to higher entropies, the temperature rises more quickly in the simulation with the new EOS. The higher temperatures then enhance electron captures, resulting in lower electron fractions and even higher temperatures as the collapse proceeds. 
As a result, the mass fraction of heavy nuclei continues to be lower at later times in the collapsing phase, where $Y_e$ becomes much lower with the new EOS than with the FYSS EOS in spite of the fact that the new EOS tends to give larger mass fractions of heavy nuclei in neutron-rich matter. The reduction of electron fraction implies a smaller mass of the inner core.

The lower electron fractions obtained in the simulation with the new EOS imply that the mass of the inner core is smaller. On the other hand, the inner core is more compact for the new EOS,  with the central density being higher. This is due to the fact that the new EOS is softer than the FYSS EOS around nuclear density. These two effects, i.e.,  the smaller mass and the higher density of the inner core tend to work oppositely for the initial strength of the shock wave generated by core bounce, with the former (latter) reducing (enhancing) the shock strength. As a consequence, the shock is a bit stronger initially for the new EOS than for the FYSS EOS. It turns out, however, that the ensuing outward propagations of the shock wave in the outer core are not much different between the two EOS's. This may be an artifact, though, 
as the
same EC rates for heavy nuclei were used in spite of different compositions in the two simulations. Indeed the deleptonization may be more suppressed for the new EOS, since neutron-rich nuclei, which have smaller EC rates in general, are more abundant in the new EOS. The results of more consistent simulations will be reported in our forthcoming paper.

Very recently, three of the authors of this paper published a paper on another EOS for supernova simulations, which we refer to as the Togashi EOS \cite{togashi17}. It is derived from
the same EOS obtained with the variational method for uniform nuclear matter by
 Togashi et al. \cite{togashi13}
and employs the SNA  based on the  Thomas-Fermi approximation.
As expected,  
the differences between our new EOS and the Togashi EOS 
are very similar to those  observed between the FYSS and  STOS EOS's  \cite{shen98a,shen98b,shen11}. 
Details are discussed in Furusawa et al. \cite{furusawa11}.
As a matter of fact, the latter two EOS's employ the same  RMF EOS 
  for uniform nuclear matter but 
 at sub-nuclear densities the former handles the ensemble of nuclei just as the new EOS does whereas the latter resorts to the SNA that the Togashi EOS adopts.
The multi-nuclear species EOS's,   i.e., the new EOS and the FYSS EOS, commonly predict smaller atomic and mass  numbers  than SNA EOS's, or the Togashi EOS and the STOS EOS \cite{burrows84,furusawa11,furusawa17c}.
In  the new EOS and the FYSS EOS, 
shell effects are taken into account and, as a result, the atomic and mass numbers grow with density in a step-wise fashion while they increase smoothly in the STOS and Togashi EOS's, in which shell effects are neglected.
Furthermore, the entropy per baryon is lower in the
latter two EOS's because they ignore the translational motion of  the representative nucleus  \cite{hempel10, furusawa11}.

Although the major differences between the new EOS and the Togashi EOS are mostly ascribed to the SNA adopted in the latter, there are certainly other differences. In fact, 
the differences between the STOS and Togashi's EOS's
 are not always consistent with those between the FYSS EOS  and the new EOS's.
For instance, the Togashi EOS predicts larger mass numbers than the STOS EOS for neutron-rich conditions ($Y_p \sim 0.1$)
 because of the larger saturation density in the former, 
which leads to greater surface energy described by 
gradient terms in the energy functional employed 
 in the Thomas-Fermi calculation. 
On the other hand, the new EOS gives smaller mass numbers than the FYSS EOS
 more often than not although there is no general trend in part due to the shell effects.
This happens because we use the same coefficient for surface tensions, $\sigma_0$,
in the new EOS and the FYSS EOS, which was originally adopted in the LDM \cite{lattimer91}.
Then, the surface energy becomes  smaller in the new EOS with the larger  saturation density.
It is true that the surface tension should be treated in a consistent manner with the EOS for uniform nuclear matter but the difference that the surface energy would make is far smaller than that stemming from the SNA approximation discussed noted 
 in the previous paragraph.

Note that the mass models of heavy and light nuclei employed in the new EOS still have some room for improvement, since there are uncertainties in our phenomenological treatment of in-medium effects. In addition, 
in the new EOS  we have adopted the same parameters in the formulae for the shell and surface energies of heavy nuclei  as well as for the Pauli energies of light nuclei
as those employed in the FYSS EOS.
They should be changed according to the theory used to describe uniform nuclear matter.
The light clusters, especially deuterons, in the new EOS
may have to be described in the framework of the variational method by using realistic nuclear forces. 
Nevertheless, we believe that the new EOS is one of the most elaborate EOS's 
constructed so far  for supernova simulations.
The tabulated  new EOS 
 is available in the public domain $^{\rm \footnotemark[1]}$.

 \footnotetext[1]{http://user.numazu-ct.ac.jp/\~{}sumi/eos/}

\section*{Acknowledgments}
S. F.  and H. N. were supported by Japan Society for the Promotion of
Science Postdoctoral Fellowships for Research Abroad.
Some numerical calculations were carried out on the PC cluster at the Center
for Computational Astrophysics, National Astronomical Observatory of Japan.
This work was supported by the RIKEN iTHES Project and  in part by the usage of supercomputer systems
through the Large Scale Simulation Program
(Nos. 15/16/-08,16/17-11) of High Energy Accelerator Research Organization (KEK)
and
Post-K Projects  (hp 150225, hp160071, hp160211,hp170031,hp170230) at K-computer, RIKEN AICS
as well as the computational resources provided by
RCNP at Osaka University, YITP at Kyoto University, University of Tokyo
and JLDG.
This work was
supported by a Grant-in-Aid for the Scientific Research
from the Ministry of Education, Culture, Sports, Science
and Technology (MEXT), Japan (24103006, 24244036, 16H03986, 15K05093, 24105008,  25400275, 26104006).

\newpage

\bibliographystyle{iopart-num.bst}
\bibliography{reference170612}
\newpage

\begin{table}[t]
\begin{tabular}{|c||c|c|c|c|c|}
\hline
   model  &  $n_{s0}$ (fm$^{-3}$)  & $E_0$ (MeV) & $K$ (MeV) & $J$ (MeV)  & $L$ (MeV) \\
 \hline
 \hline
 RMF TM1  &  0.145& -16.3  &  281   & 36.9 & 110.8 \\
 \hline
 Variational method & 0.160 &  -16.0  &  245   & 30.0 & 35.0  \\
\hline
\end{tabular}
\caption{\label{tab1_bulk}%
Bulk properties  of nuclear matter obtained in the two EOS's for uniform nuclear matter. See the beginning of  Sec.~\ref{sec:res}.}  
\end{table}

\end{document}